\documentclass[usenatbib]{mn2e}

\usepackage{txfonts}
\usepackage{graphicx}
\usepackage{natbib}

\def\nat{Nature}
\def\apj{ApJ}
\def\mnras{MNRAS}

\def\aap{A\&A}                   
\def\aapr{A\&A Rev}          
\def\apjs{ApJS}                  
\def\apjl{ApJ}                   

\def\aj{AJ}
\def\rmxaa{Rev. Mex. Astron. Astrofis}
\def\Msun{M$_{\odot}$}

\begin{document}

\author[E.~Kuiper et al.]{E.~Kuiper,$^1$\thanks{E--mail: kuiper@strw.leidenuniv.nl} N.~A.~Hatch,$^{2}$ G.~K.~Miley,$^1$ N.~P.~H.~ Nesvadba,$^3$ H.~J.~A.~R\"{o}ttgering,$^1$ 
\newauthor J.~D.~Kurk,$^4$ M.~D.~Lehnert,$^5$ R.~A.~Overzier,$^6$ L.~Pentericci,$^7$ J.~Schaye,${^1}$ B.~P.~Venemans$^8$
\\
$^1$ Leiden Observatory, Leiden University, P.O. 9513, Leiden 2300 RA, the Netherlands \\
$^2$ School of Physics and Astronomy, The University of Nottingham, University Park, Nottingham NG7 2RD, UK \\
$^3$ Institut d'Astrophysique Spatiale, Universit\'{e} Paris Sud 11, Orsay, France\\
$^4$ Max--Planck--Institut f\"{u}r Extraterrestrische Physik, Giessenbachstrasse, D-85741 Garching, Germany \\
$^5$ GEPI, Observatoire de Paris, CNRS, Universit\'{e} Denis Diderot, Meudon, France \\
$^6$ Max--Planck--Institut f\"{u}r Astrophysik, Karl--Schwarzschild Strasse 1, D--85741 Garching, Germany \\
$^7$ INAF, Osservatorio Astronomica di Roma, Via Frascati 33, 00040 Monteporzio, Italy \\
$^8$ European Southern Observatory, Karl--Schwarzschild Strasse, 85748 Garching bei M\"{unchen}, Germany \\
}

\title[Kinematics of a $z\sim2$ protocluster core]{A SINFONI view of flies in the Spiderweb: a galaxy cluster in the making}

\maketitle

\begin{abstract}
The environment of the high-$z$ radio galaxy PKS 1138-262 at $z\sim2.2$ is a prime example of a forming galaxy cluster. We use deep SINFONI integral field spectroscopy to perform a detailed study of the kinematics of the galaxies within 60~kpc of the radio core and we link this to the kinematics of the protocluster on the megaparsec scale. Identification of optical emission lines shows that 11 galaxies are at the redshift of the protocluster. The density of line emitters is more than an order of magnitude higher in the core of the protocluster with respect to the larger scale environment. This implies a galaxy overdensity in the core of $\delta_{\rm g}\sim200$ and a matter overdensity of $\delta_{\rm m}\sim70$, the latter of which is similar to the outskirts of local galaxy clusters. The velocity distribution of the confirmed satellite galaxies shows a broad, double-peaked velocity structure with $\sigma=1360\pm206$~km~s$^{-1}$. A similar broad, double-peaked distribution was found in a previous study targeting the large scale protocluster structure, indicating that a common process is acting on both small and large scales. Including all spectroscopically confirmed protocluster galaxies, a velocity dispersion of $1013\pm87$~km~s$^{-1}$ is found. We show that the protocluster has likely decoupled from the Hubble flow and is a dynamically evolved structure. Comparison to the Millenium simulation indicates that the protocluster velocity distribution is consistent with that of the most massive haloes at $z\sim2$, but we rule out that the protocluster is a fully virialized structure based on dynamical arguments and its X-ray luminosity. Comparison to merging haloes in the Millennium simulation shows that the structure as observed in and around the Spiderweb galaxy is best interpreted as being the result of a merger between two massive haloes. We propose that the merger of two subclusters can result in an increase in star formation and AGN activity in the protocluster core, therefore possibly being an important stage in the evolution of massive cD galaxies.
\end{abstract}

\section{Introduction} \label{sec:intro}

Galaxy clusters are the densest large scale environments in the known Universe and are therefore excellent laboratories for studying several of the key questions in present day astronomy. The morphology-density relation \citep[e.g.][]{dressler1980} observed in local galaxy clusters indicates that the environment of galaxies influences galaxy evolution, but when and how this happens is still unknown. Also, local galaxy clusters harbour cD galaxies, the most massive known galaxies in the Universe. Since these galaxies are exclusively located in the centers of galaxy clusters, it is likely that the cluster environment is pivotal in their formation. Finally, the emergence of large scale structure puts a strong constraint on cosmological models and parameters.

To fully understand the role of galaxy clusters in these issues, it is essential to study galaxy clusters across cosmic time. Recent years have seen the discovery of a few galaxy clusters $z>1.5$ \citep{wilson2008,papovich2010,tanaka2010,henry2010,gobat2010}, but these structures remain elusive and difficult to find at such high redshifts. One of the few methods of locating galaxy clusters at $z>2$ is targeting the environment of high-$z$ radio galaxies \citep[hereafter HzRGs,][]{miley2008}. These HzRGs show powerful extended radio emission and have large stellar masses of 10$^{11}$ to 10$^{12}$~\Msun~\citep{roccavolmerange2004,seymour2007}. As hierarchical galaxy formation dictates that the most massive galaxies originate in the densest environments, it is likely that these HzRGs are at the centres of overdensities. These overdensities may in turn be the progenitors of massive galaxy clusters. In recent years many studies have focused on finding galaxy overdensities around HzRGs \citep[e.g.][]{pascarelle1996,knopp1997,pentericci2000,venemans2005,overzier2006,overzier2008,kuiper2010,galametz2010,hatch2010}.

\begin{figure}
\resizebox{\hsize}{!}{\includegraphics{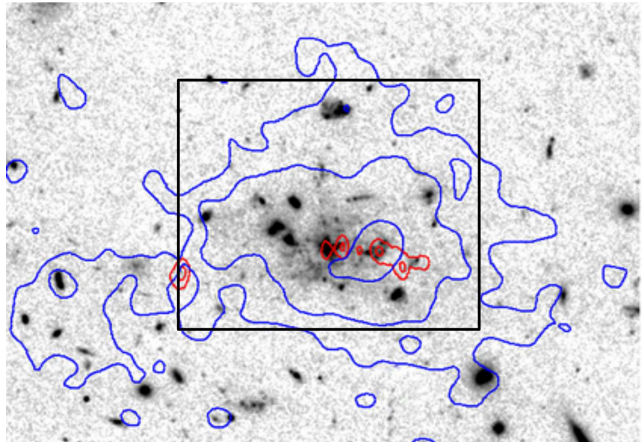}}
\caption{\label{fig:spiweb} A composite ACS ($g_{\rm 475}$+$I_{\rm 814}$) image of a $275\times200$~kpc$^{2}$ field centered on the radio galaxy PKS~1138-262. The blue and red contours, respectively, indicate the extent of the Ly$\alpha$ emission line halo and the location of non-thermal radio emission in the 8~GHz band caused by a jet \citep{pentericci1997}. The black rectangle shows the approximate outline of the field as covered by SINFONI. }
\end{figure}

One of the most studied HzRGs is PKS~1138-262 at $z\sim2.15$ (see Fig.~\ref{fig:spiweb}). The stellar mass of this radio galaxy is estimated to be $\sim10^{12}$~\Msun~\citep[][hereafter H09]{seymour2007,hatch2009}, among the largest known at $z>2$ and similar to the stellar masses found for local cD galaxies. It is surrounded by a giant Ly$\alpha$ halo powered by the AGN and young, hot stars and it is embedded in dense hot ionized gas \citep[RM$\sim6200$~rad~m$^{-2}$,][]{pentericci1997,carilli1997}. Furthermore, high resolution VLA radio observations show the presence of a radio jet with a bend. This all implies that this radio galaxy sits at the centre of a dense cluster like medium with possibly a cooling flow \citep{pentericci1997}.

Deep HST imaging shows tens of satellite galaxies, many of which are thought to be merging with the central galaxy \citep[][]{miley2006}. The restframe FUV continuum morphology of the radio galaxy is clumpy and disturbed, further strengthening the notion of strong active merging and has earned it the name of `Spiderweb Galaxy'. Such a complex morphology agrees qualitatively with predictions of hierarchical galaxy formation models \citep[e.g.][]{saro2009}.

Surrounding the central HzRG are overdensities of Ly$\alpha$ and H$\alpha$ emitting galaxies, extremely red objects (EROs), X-ray emitters and sub-mm bright galaxies \citep{pentericci2000,kurk2004a,kurk2004b,stevens2003,croft2005,zirm2008,kodama2007}. Furthermore, \citet{kurk2004b} has shown that the galaxies are spatially segregated, with the H$\alpha$ emitting galaxies and EROs being more centrally concentrated than the Ly$\alpha$ emitting galaxies. H09 showed that if the many nearby satellites are truly located in the protocluster, then a fraction may merge with the radio galaxy before $z=0$. Tidal stripping of these satellites could in turn lead to a substantial extended stellar halo as seen in local cD galaxies. Also, \citet{hatch2008} provide evidence for in-situ star formation between the individual clumps, indicating another possible method for forming such an extended stellar halo. All these aspects make the Spiderweb system a unique laboratory for studying not only important ingredients of massive galaxy formation, such as merging, downsizing and the effect of AGN feedback, but also the formation of galaxy clusters and the influence of the protocluster environment on galaxy evolution.

In this work we present results obtained using deep integral field spectroscopy data of the core of the Spiderweb protocluster\footnote[1]{There is no evidence that the radio galaxy is truly at the centre of the structure. However, the radio galaxy is a viable cD galaxy progenitor and the density of protocluster candidates around it is large. Therefore, for the sake of brevity we refer to the SINFONI field as the 'core' of the protocluster.}. Previous work on this particular region using integral field data has been done by \citet{nesvadba2006} (hereafter N06). The N06 study focused on the central radio core and its host galaxy and found evidence for the presence of strong outflows with velocities of the order of $\sim2000$~km~s$^{-1}$. These outflows are consistent with being powered by the AGN, indicating that AGN feedback plays an important role in expelling gas from galaxies thus truncating star formation.

In this follow-up study we focus on the immediate environment of the radio core and the satellite galaxies located within 60~kpc of it. We combine the integral field data with all available spectroscopic redshifts in the literature to obtain the best census of the cluster population to date. We also investigate the nature of such a protocluster structure by comparing our results to simulations. An indepth study of the internal dynamics of the brightest individual satellites will be presented in an upcoming work (Kuiper et al. 2011, in prep.). Throughout this paper we use a standard $\Lambda$CDM cosmology with $H_{\rm 0}=71$~km s$^{-1}$~Mpc$^{-1}$, $\Omega_{\rm M}=0.27$ and $\Omega_{\Lambda}=0.73$.

\section{Data} \label{sec:data}

We observed the Spiderweb Galaxy ($\alpha=11:40:48.3$, $\delta=-26:29:08.7$) with the Spectrograph for INtegral Field Observations in the Near Infrared \citep[SINFONI,][]{eisenhower2003} in seeing limited mode on UT4 at the Very Large Telescope (VLT) on several nights in December 2007 and February 2008. SINFONI is a medium-resolution, image-slicing integral-field spectrograph that has a 8\arcsec$\times$8\arcsec~field of view and a spectral resolution of approximately $R=2000-4000$ depending on the band.

The field was observed in the $J$, $H$ and $K$ bands. Based on previous work \citep{pentericci2000,kurk2004b} the redshift of the protocluster is established to be $z\sim2.15$. Therefore, the $J$ band covers the [O{\sc ii}]$\lambda3726,3729$ doublet, the $H$ band contains the [O{\sc iii}]$\lambda4959,5007$ and H$\beta$ emission lines and H$\alpha$, [N{\sc ii}]$\lambda6548,6584$ and [S{\sc ii}]$\lambda6719,6730$ are redshifted into the $K$ band. The $H$ band was given more integration time as the blue star forming satellite galaxies are likely to show [O{\sc iii}] emission. Furthermore, this line is least likely to be contaminated by neighbouring lines making it the most reliable kinematic tracer. 

\begin{table*}
\caption{\label{table1} Details of the observations. Values for the seeing are measured from the standard star observations, with the uncertainties given by the rms of individual measurements. The difference between the seeing values for $\alpha$ and $\delta$ is a natural consequence of the SINFONI image slicer being in the light path.}
\begin{tabular}{c|c|c|c|c|c}
\hline
Band  & Exp. time (sec.) & Seeing in $\alpha$ and $\delta$ (\arcsec)& Coverage & Dispersion (\AA/pixel) & Spectral Resolution ($\Delta\lambda/\lambda$)\\
\hline
\hline
$J$  & 16200 & $0.9\pm0.3$, $0.7\pm0.4$ & 13\arcsec$\times$12.5\arcsec  & 1.5 & 2000\\
$H$  & 45600 & $0.9\pm0.3$, $0.7\pm0.3$ & 16.5\arcsec$\times$16.5\arcsec  & 2.0 & 3000\\
$K$  & 28800 & $0.9\pm0.2$, $0.8\pm0.3$ & 15\arcsec$\times$12.5\arcsec  & 2.5 & 4000\\
\hline
\end{tabular}
\end{table*}

A special dithering pattern was adopted to obtain a wide field of view around the central radio galaxy, leading to an effective coverage of approximately 15\arcsec$\times$15\arcsec~centred on the radio core. Details on the observations for the various bands can be found in Table~\ref{table1}. 

Details of the data reduction can be found in N06 and \citet{nesvadba2008}, but a brief summary is given here. The data are dark subtracted and flatfielded. Curvature is measured and removed using an arc lamp after which the spectra are shifted to an absolute vacuum wavelength scale based on the OH lines in the data. This is done before sky subtraction to account for spectral flexure between the frames. The subsequent sky subtraction is done for each wavelength separately, with the sky frame being normalized to the average of the object frame in order to account for variations in the night sky emission. The three dimensional data are then reconstructed and spatially aligned using the telescope offsets as recorded in the header data. Before cube combination the individual cubes are corrected for telluric absorption. Flux calibration is done based on standard star observations and from the standard star light profile a FWHM spatial resolution of 0.7--0.9\arcsec is measured.

We also use deep Hubble Space Telescope (HST) data to supplement the SINFONI data. These data were obtained with the Advanced Camera for Surveys \citep[ACS,][]{ford1998} in the $g_{\rm 475}$ and $I_{\rm 814}$ bands \citep{miley2006} and with the Near Infrared Camera and Multi-Object Spectrometer (NICMOS) in the $J_{\rm 110}$ and $H_{\rm 160}$ bands \citep{zirm2008}.

\section{Results} \label{sec:results}

\subsection{Cluster membership} \label{sec:flies}

\begin{figure*}
\resizebox{\hsize}{!}{\includegraphics{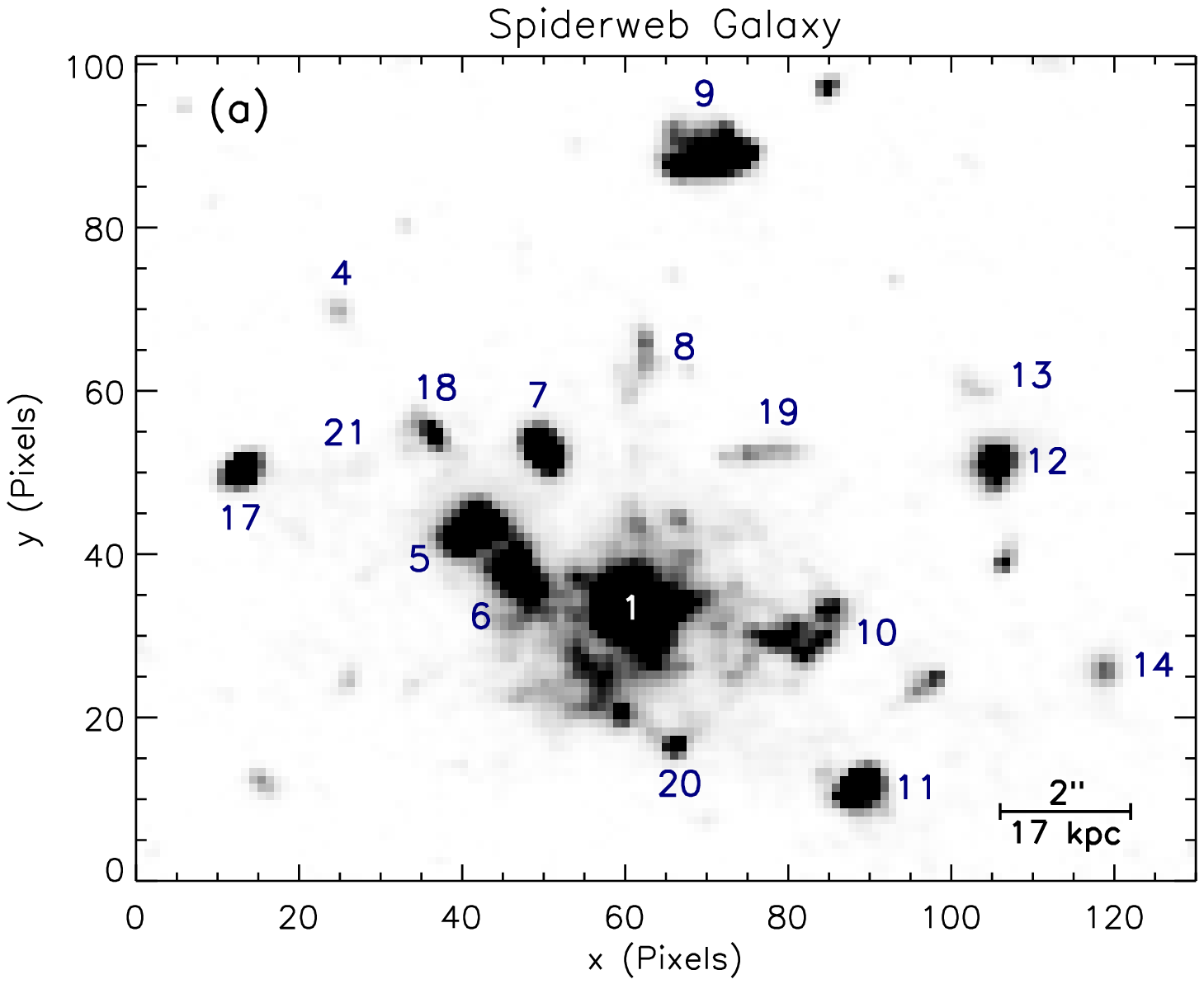}\includegraphics{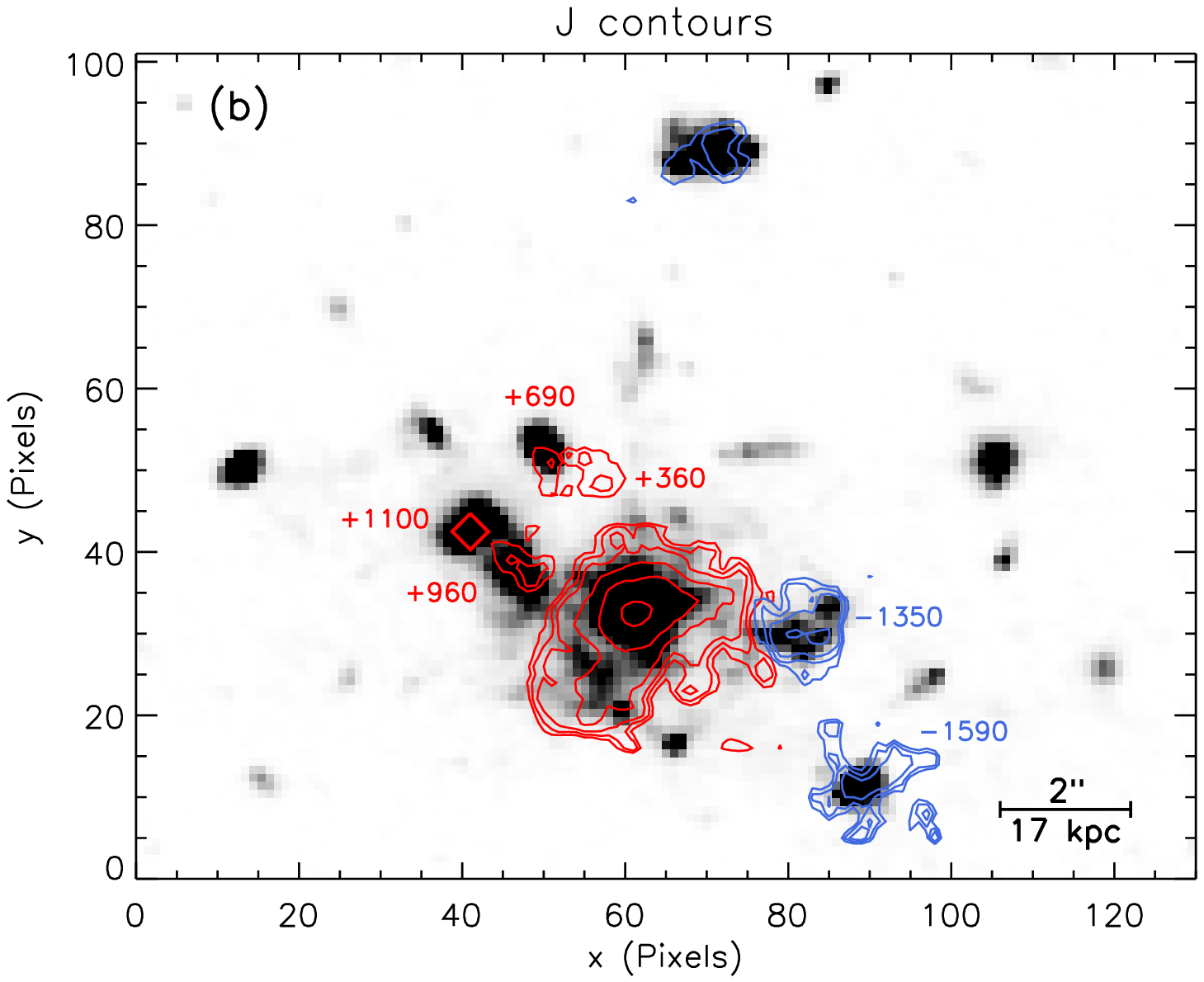}}
\resizebox{\hsize}{!}{\includegraphics{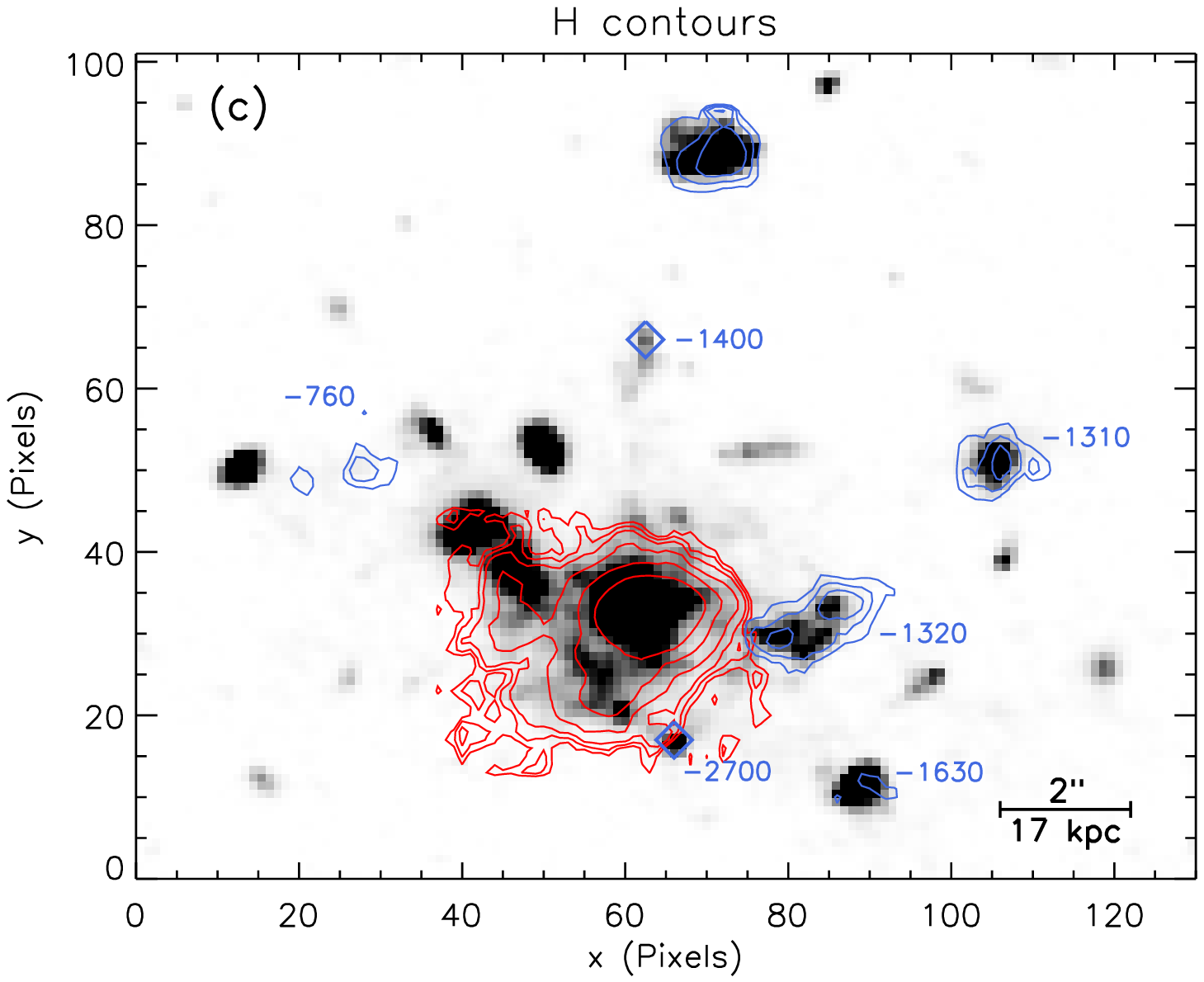}\includegraphics{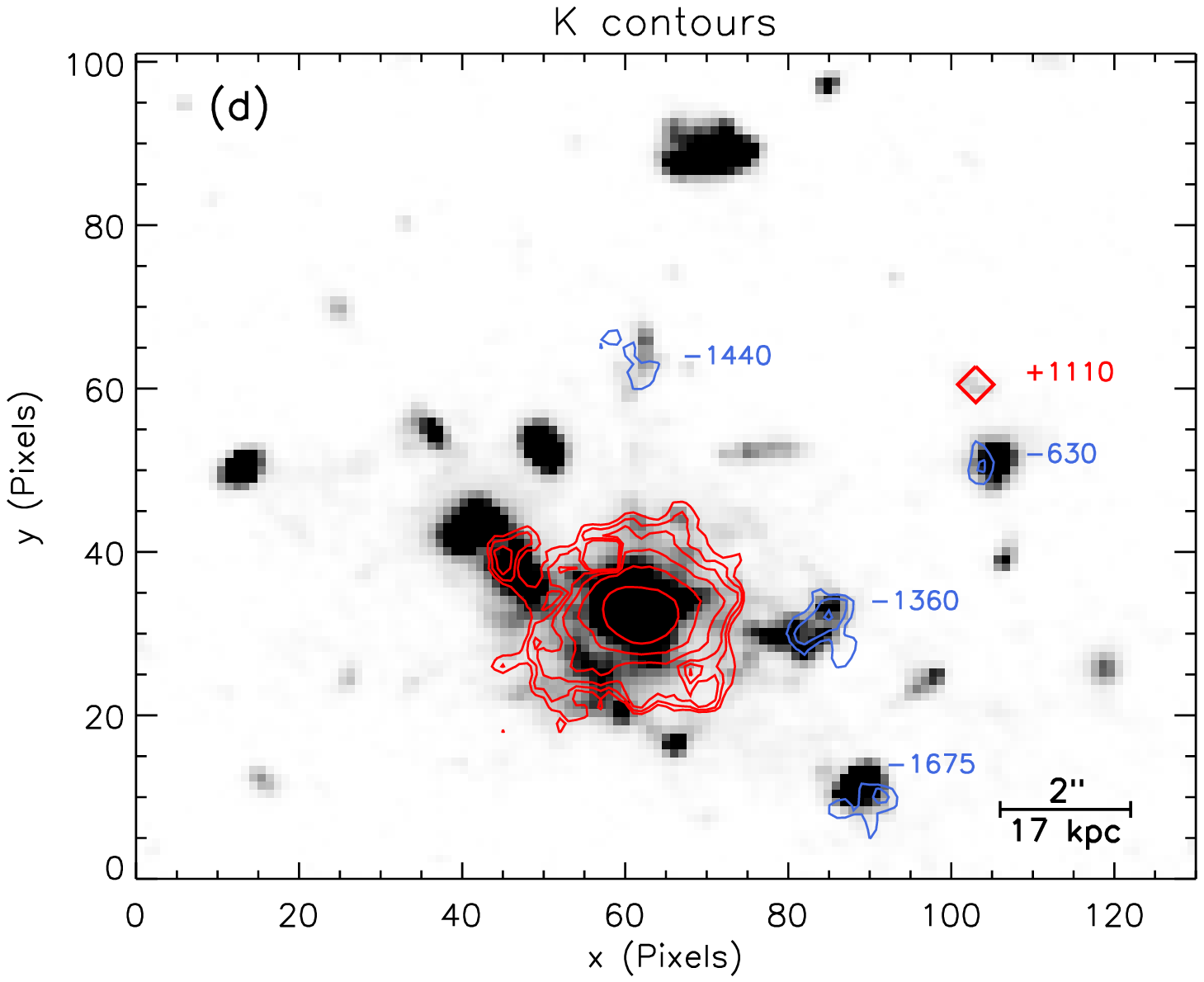}}
\caption{\label{fig:1138map} All panels show a combined $gIJH$ image of the field as covered by SINFONI. The pixel scale of the HST images has been matched to the pixel scale of the SINFONI data (0.125\arcsec~pixel$^{-1}$). In panel {\bf a} the numbers follow the labeling of the galaxies used by H09 which is also used in this work. The contours in panels {\bf b}, {\bf c} and {\bf d} show the locations of line emission in the $J$, $H$ and $K$ band, respectively. The colours of the contours indicate whether the emission is blue- or redshifted with respect to the radio galaxy. Each set of contours is obtained by summing over a narrow spectral window where line emission can be found. The line emission in question for the protocluster galaxies is [O{\sc ii}] emission in $J$ band, [O{\sc iii}] emission in $H$ band and H$\alpha$ $K$ band. The outermost and innermost contours indicate flux levels of $2.5\times10^{-19}$ and $45\times10^{-19}$~erg~s$^{-1}$~cm$^{-2}$~\AA$^{-1}$, respectively. Diamond symbols indicate objects that are too faint to yield proper contours, but do show line emission in summed spectra. The velocity offset in km~s$^{-1}$ with respect to the central radio galaxy ($z_{\rm syst}=2.1585$) is given for all objects that have line emission consistent with $z\sim2.1-2.2$.}
\end{figure*}

In each of the panels of Fig.~\ref{fig:1138map} the field covered by SINFONI is shown. The greyscale image is the sum of the $g_{\rm 475}$, $I_{\rm 814}$, $J_{\rm 110}$ and $H_{\rm 160}$ images obtained with ACS and NICMOS. The top--left panel shows the numbering convention for the individual satellites introduced by H09 which is adopted in this work as well. In addition to the 19 galaxies of H09 we add another object to the sample, \#20, as we find evidence for line emission at its location. We also mark the region between galaxies \#17 and \#18 as \#21. In H09 a bridge of red light was found at this location and this was interpreted as being a stream of gas or stars between two interacting galaxies. The SINFONI coverage does not include galaxies \#2, \#3, \#15 and \#16 and therefore these will not be discussed in this work. Galaxy \#14 does have SINFONI coverage in the $K$ band, but it is located at the edge of the field where data quality is poor. Therefore \#14 is also excluded from this work.

The contours in panels {\bf b}, {\bf c} and {\bf d} of Fig.~\ref{fig:1138map} indicate the locations of line emission in each of the three bands. The contours have been produced for each satellite galaxy individually. This was done by creating a cut-out at the location of the satellite and summing for each pixel in the cutout over the spectral range where evidence for line emission can be found. The resulting line image was used for calculating the contours.  

\begin{figure*}
\resizebox{\hsize}{!}{\includegraphics{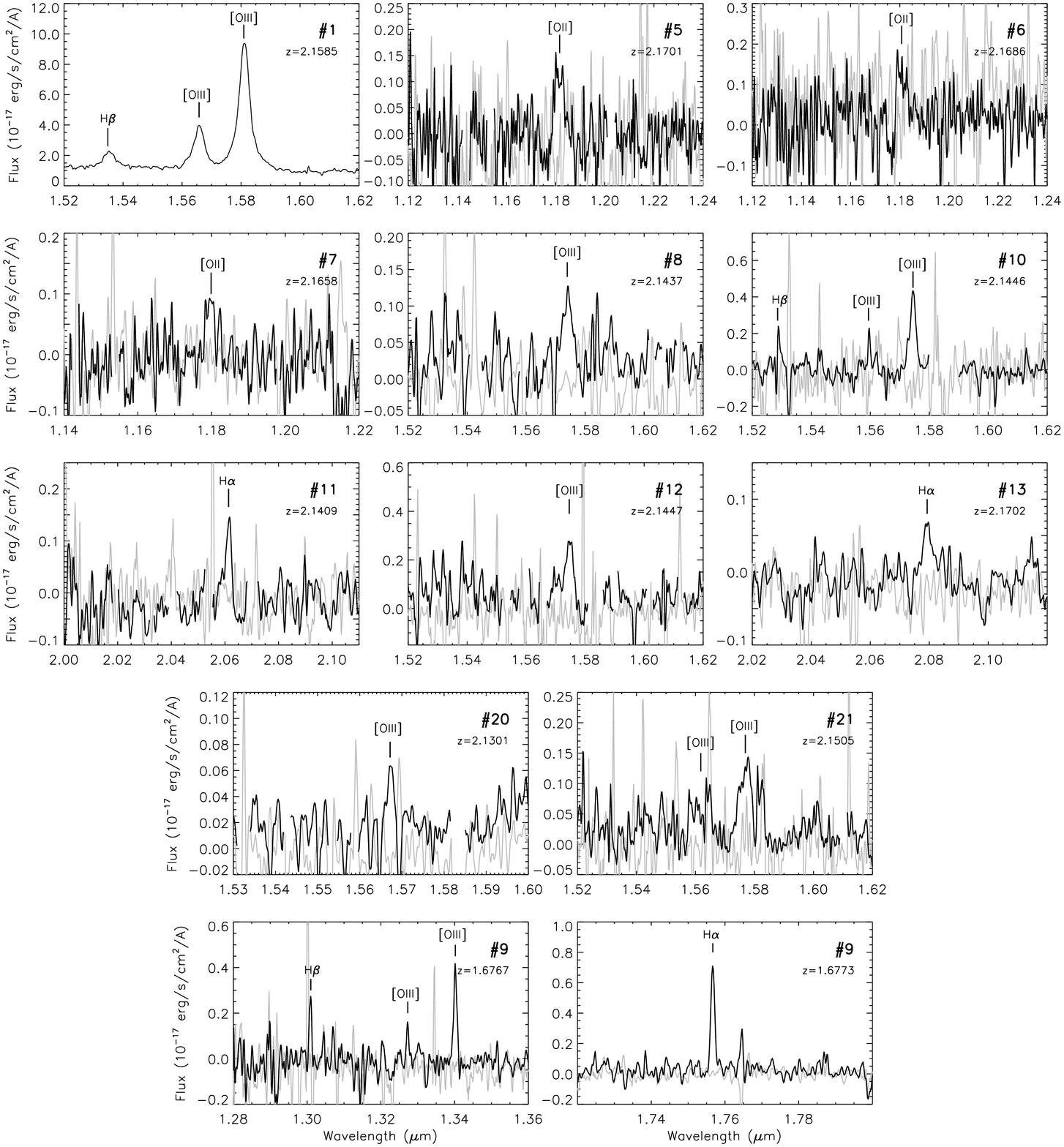}}
\caption{\label{fig:stacks} Summed spectra of the galaxies in the SINFONI field that show emission lines. Only the strongest emission lines are shown. For the satellite galaxies sky spectra are also shown in gray. The sky spectra have been extracted from locations close to the galaxies using the same apertures. For clarity, the emission lines have been labeled and patches of poor night-skyline residuals have not been plotted. Eleven galaxies are identified as protocluster members and one galaxy (\#9) is identified as a foreground galaxy.}
\end{figure*}

As can be seen in Fig.~\ref{fig:1138map}, there are multiple sources of line emission, most of which are associated with the satellite galaxies. Some of the objects that show no clear evidence of line emission in single pixels do show evidence for line emission after summing the pixels associated with continuum emission in the ACS and NICMOS data. One or more emission lines consistent with $z\sim2.15$ are detected for 11 galaxies. These galaxies are \#1, \#5, \#6, \#7, \#8, \#10, \#11, \#12, \#13, \#20 and \#21. Figure~\ref{fig:stacks} shows the strongest emission lines for each of these confirmed satellites. The spectra for each of the galaxies have been obtained by summing the spectra of individual pixels with line emission. For the faint or initially undetected objects all pixels within the seeing disk at the location of the galaxy have been summed. Sky spectra for the satellite galaxies are also shown in Fig.~\ref{fig:stacks}. These have been extracted using the same apertures as used for the individual galaxies and they give an indication of the location and severity of sky line contamination. No significant line emission is found for \#4, \#17, \#18 and \#19.

Figure~\ref{fig:stacks} also shows the spectrum of one galaxy (\#9) that is identified as being a low redshift interloper through the identification of [O{\sc iii}] and H$\alpha$ at $z=1.677$. This is surprising, as previous studies have provided ample evidence for it being at the redshift of the protocluster. The Ly$\alpha$ narrowband imaging of \citet{pentericci1997} shows a significant and distinct source of emission at the location of galaxy \#9 and subsequent spectroscopy detected an emission line that is consistent with being Ly$\alpha$ at $z\sim2.15$ \citep{kurk2003}. There are no strong emission lines at $z=1.677$ that could mimic Ly$\alpha$ at the protocluster redshift. It is therefore most likely that the emission at $\sim$3840~\AA~is Ly$\alpha$ emission from the extended Ly$\alpha$ halo surrounding the central radio galaxy rather than Ly$\alpha$ emission from the galaxy itself.

In addition to \#9, several of the other objects in the SINFONI field have been previously targeted for spectroscopy \citep{kurk2003}. Objects \#1, \#5/\#6, \#7, \#10 and \#11 have all been shown to have Ly$\alpha$ emission at $z\sim2.15$. However, a comparison between the results presented in this paper and those obtained from Ly$\alpha$ spectroscopy shows in general large differences. Five out of six objects have velocities based on Ly$\alpha$ that differ by 500~km~s$^{-1}-1500$~km~s$^{-1}$ with respect to the velocities presented in this paper. Only object \#11 shows consistent redshifts. The offsets found for the other objects do not indicate any systematic trend. This is in accordance with \citet{pentericci1997}, who found no evidence for ordered motion such as rotation. The resonant nature of the Ly$\alpha$ line can thus cause the measured redshift to deviate significantly from the true redshift. Therefore caution must be used when interpreting the redshifts obtained through spectroscopic confirmation of Ly$\alpha$ alone.

A full list of all detected emission lines and their corresponding redshifts, velocities, line widths and fluxes can be found in Table~\ref{table2}. The brightest galaxies for which emission lines are detected in individual pixels have been corrected for internal kinematic structure. This has been done by shifting the individual spectra such that the line centres in the individual pixels match the redshift of the galaxy as a whole. Uncertainties are calculated by varying the summed spectra using a normal distribution characterized by the rms noise of the spectrum in question. This is repeated 1000 times and the standard deviation of the resulting parameter distributions are taken as the $1\sigma$ uncertainties. 

\begin{table*}
\caption{\label{table2} Line properties for each of the protocluster candidates. The systemic redshift is taken to be the redshift of \#1 as measured from the [O{\sc iii}]$\lambda5007$ emission line. $^{a}$Where possible the galaxies have been corrected for internal kinematic structure. $^{b}$Values given are for H$\alpha$+[N{\sc ii}]. $^{c}$The [S{\sc ii}] doublet is unresolved. $^{d}$$3\sigma$ upper limits in $J/H/K$ band for a FWHM of 500~km~s$^{-1}$. $^{e}$Lines likely contaminated by emission from the central radio source.}
\begin{minipage}[b] {\hsize}\centering
\begin{tabular}{c|r|r|r|r}
\hline
Object & Line & $z$ & $v$~(km~s$^{-1}$) & Flux (10$^{-17}$ erg s$^{-1}$ cm$^{-2}$ \AA$^{-1}$)$^{a}$ \\
\hline
\hline
\#1/RG & [O{\sc ii}] & $2.1590\pm0.0002$ & +40 & $128\pm3$\\
 & [Ne{\sc iii}]$\lambda3869$ & $2.1595\pm0.0010$ & +10 & $35.9\pm4.7$\\
  & [Ne{\sc iii}]$\lambda3968$ & $2.1555\pm0.0015$ & +10 & $16.7\pm2.7$\\
 & H$\beta$ & $2.1590\pm0.0002$ & +47 & $50.2\pm0.9$ \\
 & [O{\sc iii}]$\lambda4959$ & $2.1584\pm0.0001$ & -15 & $137\pm1$ \\
 & [O{\sc iii}]$\lambda5007$ & $2.1585\pm0.0001$ & 0 &  $477\pm1$\\
 & H$\alpha$ & $2.1585\pm0.0001$ & -5 & $582\pm4^{b}$\\
 & [S{\sc ii}]$^{c}$ & $2.1590\pm0.0018$&  +40 & $149\pm4$\\
\#4 & - & - & -  & $<2.5/0.6/0.7^{d}$ \\
\#5 & [O{\sc ii}] & $2.1701\pm0.0016$ & +1100 & $5.6\pm0.8$ \\
 & [O{\sc iii}] & $2.1695\pm0.0004$ & +1040 & $5.9\pm0.7$ \\
 & H$\alpha$ & $2.1699\pm0.0015$ & +980 & $4.3\pm0.7$\\
\#6 & [O{\sc ii}] & $2.1686\pm0.0018$ & +950 & $5.8\pm1.0$\\
 & [O{\sc iii}] & $2.1683\pm0.0002$ & +925 & $18.1\pm1.1$$^{e}$\\
 & H$\alpha$ & $2.1706\pm0.0005$ & +1220 & $35.8\pm2.2$$^{e}$ \\
\#7 & [O{\sc ii}] & $2.1658\pm0.0011$& +690 & $3.2\pm0.9$ \\
\#8 & [O{\sc iii}] & $2.1437\pm0.0007$& -1410 & $4.0\pm0.7$ \\
 & H$\alpha$ & $2.1434\pm0.0002$ & -1440 & $5.5\pm0.9$ \\
\#9 & [O{\sc iii}] & $1.6767\pm0.0005$ & - & $4.6\pm0.6$ \\
 & H$\alpha$ & $1.6773\pm0.0001$ & - & $9.1\pm0.6$\\
\#10 & [O{\sc ii}] & $2.1443\pm0.0006$ & -1350 & $11.6\pm1.4$ \\
& [O{\sc iii}] & $2.1446\pm0.0001$ & -1325 & $11.1\pm1.0$ \\
& H$\alpha$ & $2.1442\pm0.0002$ & -1360 & $10.4\pm1.0$ \\
\#11 & [O{\sc ii}] & $2.1418\pm0.0004$ & -1590 & $6.4\pm1.2$ \\
 & [O{\sc iii}] & $2.1415\pm0.0008$ & -1620 & $2.6\pm0.7$ \\
 & H$\alpha$ & $2.1409\pm0.0004$ & -1675 & $3.9\pm0.6$ \\
\#12 & [O{\sc iii}] & $2.1447\pm0.0023$ & -1315 & $8.2\pm1.5$ \\
\#13 & H$\alpha$ & $2.1702\pm0.0011$ & +1110 &  $2.6\pm0.4$ \\
\#17 & - & - & - & $<1.7/0.6/1.0^{d}$ \\ 
\#18 & - & - & - & $<1.1/0.6/0.7^{d}$ \\ 
\#19 & - & - & - & $<0.7/0.4/0.6^{d}$ \\
\#20 & [O{\sc iii}] & $2.1301\pm0.0010$ & -2700 & $0.9\pm0.3$\\
\#21 & [O{\sc iii}] & $2.1505\pm0.0010$ & -765 & $6.1\pm0.7$\\
\hline
\end{tabular}
\end{minipage}
\end{table*}

\subsection{Overdensity}

The confirmation of 11 protocluster galaxies within a $\sim60$~kpc radius makes this field extraordinarily dense. The surface number density of the core region is $1.8\times10^{2}$~arcmin$^{-2}$ or $7\times10^{-4}$~kpc$^{-2}$ in physical units. This is likely a lower limit to the actual value as quiescent galaxies without line emission cannot be spectroscopically confirmed with the SINFONI data. 

To illustrate the extreme denseness of the region around the Spiderweb galaxy we compare the density in the SINFONI field to that of the larger protocluster field. \citet{kurk2000} found a total of 50 Ly$\alpha$ emitter candidates in a 35.4~arcmin$^{2}$ field centered on the radio galaxy. In \citet{kurk2004b} (hereafter K04) a sample of 40 candidate H$\alpha$ emitters was identified within a field of $\sim12$~arcmin$^{2}$. Respectively, six and three\footnote[2]{\#5, \#6 and \#10 were taken as part of \#1 in K04 and therefore not identified as individual H$\alpha$ emitters. \#8 and \#13 were too faint to be included in the H$\alpha$ emitter sample.} of these line emitters are located in the SINFONI field of 0.0756~arcmin$^{2}$. Thus, it is found that the SINFONI field is $\sim56\pm10$ times denser than the protocluster as a whole for the Ly$\alpha$ emitters and $\sim12\pm4$ times as dense for the H$\alpha$ emitters. Here the uncertainties are determined using Poisson statistics. Since the large scale protocluster field is respectively $4\pm2$ and $14\pm2$ times denser than the field at $z\sim2.2$ \citep[][]{kurk2004a,hatch2011} this indicates that the protocluster core has a galaxy overdensity of $\sim200$. 

The bias factor of star forming galaxies at high redshift is typically 1--5 \citep[e.g.][]{adelberger1998,giavalisco1998,ouchi2004}. Thus for our $z\sim2$ star forming galaxies we assume a bias factor of 3, which is roughly consistent with the trend observed in \citet{marinoni2005}. This implies a matter overdensity in the core of $\sim70$. Local galaxy clusters have matter overdensities at their virial radii of $\sim100$. The matter overdensity in the core of the Spiderweb protocluster is thus similar to what is observed in the outskirts of local galaxy clusters. 

Such core overdensities are rare, but not unique, among high-$z$ protoclusters. Out of the sample of eight protoclusters studied in \citet{venemans2007} none have Ly$\alpha$ emitters within 60~kpc of the radio galaxy\footnote[3]{Here the assumption is made that the radio galaxy traces the core of the protocluster}. For the H-alpha emitters only two other radio-galaxy protoclusters have been imaged. \citet{itanaka2010} found an overdensity of $\delta_{\rm g}=4\pm2$ of H$\alpha$ emitters around 4C23.56 ($z\sim2.48$), but none were found in direct vicinity of the radio galaxy. In another study by \citet{hatch2011} an overdensity of $\delta_{\rm g}=12\pm2$ was found in the field of radio galaxy 4C10.48 at $z\sim2.35$. A significant number of these was found close to the radio galaxy. The spatial distribution, however, shows a striking alignment of the galaxies with the radio jet indicating that these objects may be experiencing jet-induced star formation. It is thus not clear whether the origin of the overdensity in 4C10.48 is of the same nature as that seen around the Spiderweb galaxy.

\subsection{Velocity distribution} \label{sec:velodis}

We have obtained redshifts for a large number of objects in the field. Before, protocluster candidacy was determined using Ly$\alpha$ narrowband imaging and follow-up spectroscopy, but as we have shown for satellite \#9 in Sect.~\ref{sec:flies}, the presence of the extended Ly$\alpha$ halo can lead to incorrect redshifts. No extended optical line emission is seen at distances larger than $\sim15$~kpc from the radio core and thus these emission lines are likely to originate from the satellite galaxies.

\begin{figure}
\resizebox{\hsize}{!}{\includegraphics{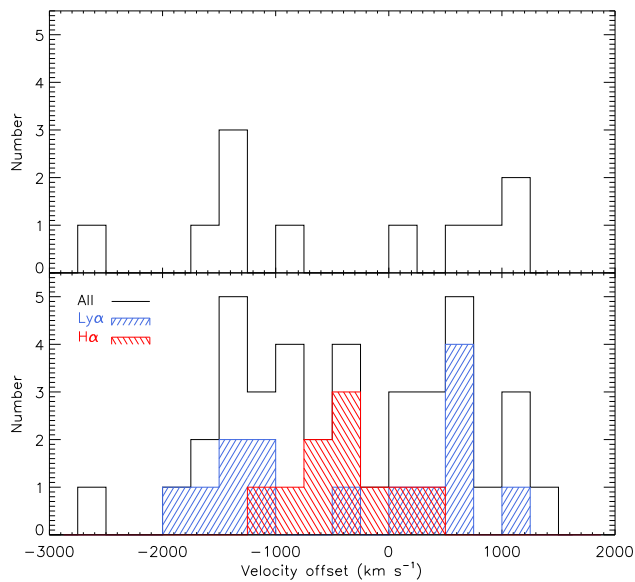}}
\caption{\label{fig:velohist} Top panel: velocity distribution of the confirmed protocluster members found in this work. Velocities are with respect to $z_{\rm [OIII]}$ of the radio core (\#1). Two groups can be identified that are blue- and redshifted with respect to the radio galaxy. Bottom panel: velocity distribution of the full Spiderweb sample. The velocity distributions of P00 and K04 are indicated by the blue and red hatched regions, respectively.}
\end{figure}

The velocity distribution of the spectroscopically confirmed protocluster core galaxies is shown in the top panel of Fig.~\ref{fig:velohist}. Velocities are taken with respect to the systemic redshift of the radio galaxy. The detection of stellar absorption features would unequivocally set the redshift of the stellar content of the radio galaxy. However, none are found and therefore we rely on emission lines. As shown in N06, the internal velocity distribution of the central radio source based on [O{\sc iii}] covers a large range of velocities of $\sim1500$~km~s$^{-1}$ ($2.158<z<2.173$). This indicates the presence of strong outflows. It is therefore not trivial to assign a systemic redshift to this object. In this work the systemic redshift ($z=2.1585$) is chosen to be the redshift of the location of the HST continuum emission (zone 3 in N06). It may be that this region hosts an outflow that is blueshifted with respect to the stellar content of the galaxy. This means the actual redshift of the central radio source may be significantly higher; up to $z\sim2.170$. This would result in a shift  of the zeropoint in Fig.~\ref{fig:velohist} of several 100~km~s$^{-1}$ towards larger positive velocities. The shape of the velocity distribution, however, will remain the same. 

In Fig.~\ref{fig:velohist} a striking subclustering in the velocity distribution can be seen with two subgroups at $\Delta v\sim-1500$km~s$^{-1}$ and $\Delta v\sim+900$km~s$^{-1}$. The large separation between the subgroups implies a large velocity dispersion of the satellite galaxies as a whole. Assuming that the underlying distribution is a single Gaussian, a value of $1360\pm206$~km~s$^{-1}$ is found using the Gapper scale estimator \citep{beers1990}.

A comparison of the velocity distribution to the results of previous studies shows both resemblances and differences. \citet[][P00]{pentericci2000} used the Ly$\alpha$ emission line to spectroscopically confirm 14 objects as being at the protocluster redshift. All 14 objects are located outside the SINFONI field of view. They thus trace the large scale protocluster structure and contamination from the extended Ly$\alpha$ halo is unlikely. Another study by K04 obtained spectroscopy on candidate H$\alpha$ emitters in the Spiderweb protocluster. Nine objects were confirmed to be in the protocluster. Finally, the studies of \citet{croft2005} and \citet{doherty2010} have respectively confirmed two X-ray emitting objects and two red galaxies to be part of the protocluster structure. The resulting full velocity distribution, including all confirmed galaxies, is shown in the bottom panel of Fig.~\ref{fig:velohist}. For clarity the distributions of P00 and K04 have been indicated by blue and red hatched histograms, respectively. 

First considering the P00 distribution, we see that even though these studies are fully independent and target different scales within the protocluster, the velocity distribution of the Ly$\alpha$ emitters is qualitatively the same as in the protocluster core. The subgroups in the P00 distribution are located at $z=2.145\pm0.002$ and $z=2.164\pm0.002$ which is similar to the locations of the subgroups found in this work (see Table 2). The fact that both the large and small scale velocity distribution are broad and double-peaked indicates that there is a common process acting on both large and small scales. Therefore, the evolution of the central Spiderweb galaxy and the larger protocluster structure may be linked.  We will come back to this in Sect.~\ref{sec:core}.

In contrast to this, the K04 velocity distribution shows neither a resemblance to the velocity distribution found in this study nor to that of P00. The nine confirmed H$\alpha$ emitters have a relatively narrow velocity distribution with redshifts in the range $2.1463 < z < 2.1636$. It is not clear why the H$\alpha$ emitters show such a different velocity distribution, because the narrowband transmission curve covers the entire velocity range spanned by the protocluster galaxies. Considering the poor number statistics it may therefore be attributed to sampling. Nevertheless, the full distribution, including the velocities of all confirmed objects, is broad and yields a velocity dispersion of $1013\pm87$~km~s$^{-1}$. 

Are such high velocity dispersions common in high-$z$ protoclusters? In \citet{venemans2007} a sample of six protoclusters with redshifts in the range of $2<z<5$ was studied. Most of the protoclusters have velocity dispersions significantly smaller than 1000~km~s$^{-1}$, but the protocluster around MRC~0052-241 (0052) at $z=2.8600$ was found to have a velocity dispersion of $980\pm120$~km~s$^{-1}$. Also, the velocity distribution of 0052 shows signs of bimodality. It thus seems that although the Spiderweb protocluster is exceptional, it is not unique in showing these characteristics.

\citet{venemans2007} also found an increase in velocity dispersion with decreasing redshift. Although the Spiderweb protocluster is in line with this trend, its velocity dispersion is significantly higher than the $z\sim1.2$ galaxy cluster RDCS~1252.9--2927, which was used as a `low' redshift control case in the \citet{venemans2007} study. This indicates that even when considering this trend, the Spiderweb protocluster remains an exceptional case.

Another possible cause for the large velocity dispersion has been proposed by P00. P00 speculated that the double--peaked profile hints at a possible merger of two groups of galaxies. Several cases of merging clusters are known at low $z$ and some do show a clear segregation in velocity space \citep[e.g. Abell~1750, Abell~2744,][]{ramirez1990,owers2011}. Considering this possibility, we fit a double Gaussian to the full Spiderweb velocity distribution. We find that the two groups are separated by a velocity difference of 1600~km~s$^{-1}$ and have velocity dispersions of 492~km~s$^{-1}$ and 417~km~s$^{-1}$. 

A Kolmogorov-Smirnov (KS) test was used to investigate whether the single or double component model fits the observations better. Although formally both models are consistent with the observed distribution, the probability that the two distributions match is 0.98 for the double component model whereas is it 0.66 for the single component model. Also, a normalized tail index \citep{bird1993} yields 0.82 indicating that a uniform distribution better describes the complete distribution rather than a single Gaussian. Note that if the Spiderweb protocluster is indeed a double or merging system, then the comparisons to the results of \citet{venemans2007} discussed above will yield significantly different results.

Could the subclustering found in the core of the protocluster be an artefact of the data? The wavelength range $15800 \le \lambda \le 15850$ ($2.156 \le z \le 2.166$) in the $H$ band is one of the regions particularly affected by poor night--skyline residuals. This may lead to the galaxies with weak optical emission lines in this particular wavelength range being missed. If all of the four unconfirmed protocluster candidates are indeed in this specific range then the core distribution would more closely resemble a flat distribution. However, we argue that the velocity distribution of the complete sample will remain broad and its shape will not change significantly.

\section{Discussion} \label{sec:disc}

\subsection{A galaxy cluster progenitor} \label{sec:field}

The Spiderweb protocluster is known to harbour galaxy overdensities (see Sect.~\ref{sec:intro}) and is therefore thought to evolve into a galaxy cluster. It is however not guaranteed that an overdensity at high-$z$ will evolve into a present day massive galaxy cluster. To determine whether the Spiderweb protocluster is truly a forming galaxy cluster, we attempt to determine whether the dynamical state of the galaxies in the overdensity is significantly different from what is expected in field environments. 

\begin{figure}
\resizebox{\hsize}{!}{\includegraphics{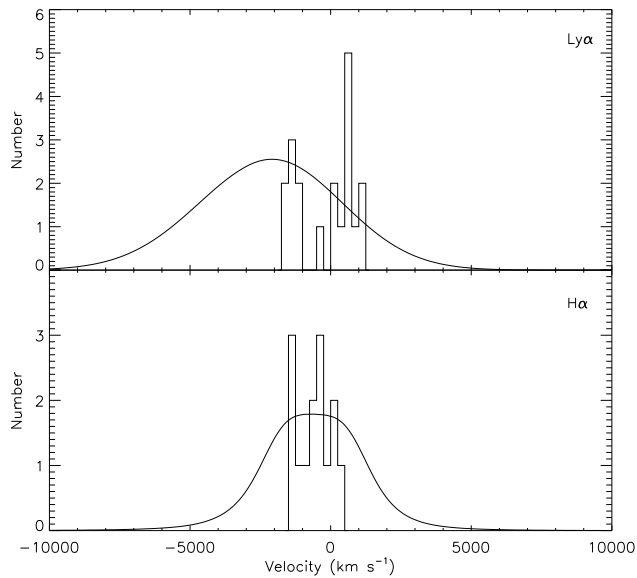}}
\caption{\label{fig:field} The velocity distributions of all spectroscopically confirmed Ly$\alpha$ and H$\alpha$ emitter candidates. Also shown are the respective transmission curves (or selection functions) as function of velocity with respect to the radio galaxy. The Ly$\alpha$ velocity distribution differs from the selection function indicating it is not a field environment.}
\end{figure}

Shown in Fig.~\ref{fig:field} are the velocity distributions of all spectroscopically confirmed galaxies that have been marked as line emitters in P00 and K04. Also plotted are the selection functions based on the transmission curves of the narrowband filters used in those studies. These selection functions should be good approximations of the expected field velocity distributions. The H$\alpha$ emitters shown in the bottom panel follow the expected distribution quite well. A KS test yields a probability of 20 per cent that the observed distribution is drawn from the expected velocity distribution. This indicates that we cannot conclude whether there is a significant difference between the two distributions. All of the 19 Ly$\alpha$ emitters, however, are located at higher velocities than the mean velocity implied by the selection function. Applying a KS test we determine that there is a $5\times10^{-6}$ probability that the Ly$\alpha$ emitters in the overdensity are drawn from the expected field velocity distribution. Thus the two distributions are different at the $\sim4.5\sigma$ level. From this we conclude that the Ly$\alpha$ emitters are associated with the radio galaxy. In fact, this is likely the case for all the galaxies considered, since the Ly$\alpha$ emitters cover the same velocity range.

These results also imply that the protocluster structure has dropped out of the Hubble flow. Assuming this is indeed the case, how does the dynamical state of the protocluster compare to that of the field if the Hubble flow is removed? To investigate this we compare to galaxies that reside in field environments at $z\sim2$ in the Millennium simulation \citep{springel2005,delucia2007}. The Millennium simulation traces cosmological evolution at a number of discrete redshift `snapshots'. The resulting field velocity distribution will therefore indicate the deviation from the Hubble flow. The redshift chosen for this purpose is $z=2.24$, which is the snapshot closest to the redshift of the actual protocluster. 

The total area probed by P00 is approximately 100 comoving Mpc$^{2}$. To obtain proper statistics, we therefore select three separate boxes with sides of 50 comoving Mpc from the Millennium simulation. Each of the $xy$, $xz$ and $yz$ planes is subsequently divided into a grid of $10\times10$~Mpc$^{2}$ fields, thus yielding a total of 75 fields per box. To ensure that no large scale structures are present, we exclude all fields with number densities that satisfy $n-\bar{n} >2\sigma$, with $\bar{n}$ the median number density. Approximately 10 per cent of the fields are excluded this way. Changing this limit does not strongly influence the final result. We also apply a magnitude cut to the restframe $R$ band magnitude of the simulated galaxies. This cut is placed arbitrarily at $R<27$ since varying the exact value of the cut between $26 < R < 30$ does not alter any of the conclusions presented here.

\begin{figure}
\resizebox{\hsize}{!}{\includegraphics{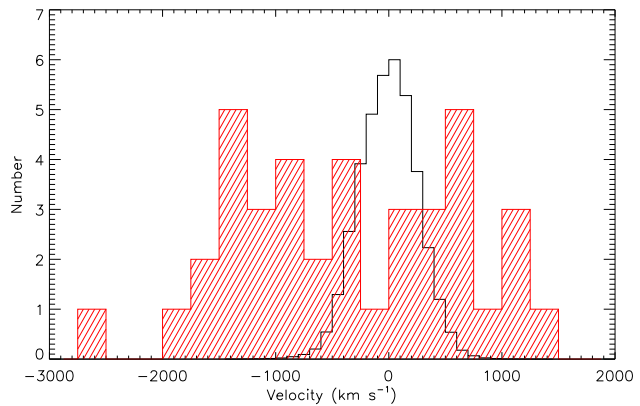}}
\caption{\label{fig:fieldmil} Composite field velocity distribution of peculiar velocities obtained from the Millennium simulation. Also shown as the hatched histogram is the Spiderweb velocity distribution. The Millennium field distribution has been scaled down to facilitate comparison. The field velocity distribution is significantly narrower than the Spiderweb velocity distribution.}
\end{figure}

A composite galaxy velocity distribution of the remaining fields is shown in Fig.~\ref{fig:fieldmil}. When compared to the protocluster distribution, it is seen that the composite field velocity distribution is relatively narrow. A KS test is used to determine whether the full Spiderweb velocity distribution is consistent with the field velocity distribution. A probability of 9$\times10^{-7}$ is found, indicating that the two distributions differ at the $\sim5\sigma$ level. Applying a KS test to each of the individual field distributions shows that all differ at least at the 3$\sigma$ level with 95 per cent of the fields differing at $>4\sigma$. Therefore, if the protocluster has indeed fully dropped out of the Hubble flow, then it differs significantly from what is expected from a field environment.

We conclude that the Spiderweb field is not only overdense, but also kinematically distinct from regions where the Hubble flow dominates. The Spiderweb protocluster is thus a dynamically evolved, collapsing (or collapsed) structure. 

\subsection{Formation scenarios} \label{sec:formscen}

With the confirmation of a total of 38 galaxies at the redshift of the radio galaxy and from the fact that this is a dynamically evolved structure, it is apparent that the Spiderweb is the progenitor of a galaxy cluster. As it is still unclear how galaxy clusters are formed, we review here two possible formation scenarios.

\subsubsection{Single virialized structure}

The first option we consider is the scenario in which the Spiderweb protocluster is a single virialized structure. This can be considered by comparing the velocity distribution presented in this paper to results obtained with the Millennium simulation. Although we know little of the dark matter halo that hosts the protocluster, the dense nature of the Spiderweb field indicates that this is likely one of the most massive dark matter haloes at $z\sim2.15$. We therefore investigate the properties of the 25 most massive dark matter haloes at $z=2.24$ in the Millennium simulation. These haloes have masses ranging from 0.75--1.4$\times10^{14}$~\Msun~and galaxy velocity dispersions in the range of 450-900~km~s$^{-1}$. Visual inspection shows that one of the haloes is in the process of merging with another massive halo. This particular halo is not included in the comparison for the single halo scenario.

Using a KS test we investigate whether the double-peaked structure found for the Spiderweb can be drawn from the velocity distributions obtained for the Millennium haloes. This is done for all three velocity components separately. Using the full Spiderweb velocity distribution we find that 19 haloes yield probabilities indicating that they are different from the protocluster distribution at at least the 2$\sigma$ level. The remaining 6 haloes yield probabilities of $\sim10$~per~cent. The possibility that the Spiderweb protocluster is a single massive structure can therefore not be wholly excluded on these grounds.

Another test for the virialized structure scenario is supplied by the protocluster mass. If virialized, it is possible to determine the mass contained within the area covered by the SINFONI field using \citep{small1998}:
\begin{equation}
M_{\rm vir}=\frac{6}{G}\sigma^{2}r_{\rm vir}
\end{equation} 
with $\sigma$ the velocity dispersion obtained in Sect.~\ref{sec:velodis} and $r_{\rm vir}$ the mean projected harmonic separation given by
\begin{equation}
r_{\rm vir}=\frac{\pi}{2}\frac{N(N-1)}{2}\left(\sum_{i<j}\frac{1}{|r_{i}-r_{j}|}\right)^{-1}.
\end{equation}
Here $N$ is the number of objects considered. Using $\sigma=1360$~km~s$^{-1}$ and finding $r_{\rm vir,core}=55$~kpc, a dynamical mass of $\sim10^{14}$~\Msun~is found for the core region of the protocluster. 

For the complete Spiderweb galaxy sample the velocity dispersion is 1013~km~s$^{-1}$ and we approximate $r_{\rm vir}$ by using only the objects of P00 as the K04 and SINFONI objects are taken from data with smaller field of view: including these would therefore bias the radius to smaller values. We find that the virial radius of the Spiderweb is 1.2~Mpc, implying a total mass of $\sim2\times10^{15}$~\Msun. 

Using the scaling relations presented in \citet{rykoff2008} and the total protocluster mass of $2\times10^{15}$~\Msun, the X-ray luminosity of the cluster gas in restframe 0.1--2.4~keV can be calculated to be $\sim1.5\times10^{45}$~erg~s$^{-1}$. Assuming $k_{\rm B}T_{\rm X}\sim 5-7$~keV we convert this to the range 2--10~keV and find $3-6\times10^{44}$~erg~s$^{-1}$. This can be compared to the upper limit on the X-ray luminosity of an extended cluster atmosphere of $1.5\times10^{44}$~erg~s$^{-1}$ presented in \citet{carilli2002}. If the protocluster is indeed virialized, it would have been detected. This has been indicated in K04 as well. Furthermore, a total mass of $\sim10^{15}$~\Msun~would rank this halo among the most massive known haloes and the probability that such a massive halo exists at this redshift is negligible according to $\Lambda$CDM cosmology \citep[e.g.][]{jee2009}. We thus conclude that the protocluster is not a virialized structure. If only the core region were virialized the expected X-ray luminosity would drop to $8\times10^{42}$~erg~s$^{-1}$. This possibility can therefore not be excluded. 

\citet{saro2009}, who presented high resolution hydrodynamical simulations of high-$z$ protoclusters modelled after the Spiderweb system, also argued that the inner region of a Spiderweb-like protocluster ($r<400$~kpc) can be virialized at $z\sim2$. However, the velocity dispersion in the simulated cluster is significantly lower than what is presented in this study, making it unclear whether it is applicable to the Spiderweb system.

\subsubsection{Merger of haloes} \label{sec:merger}

Given the broad double-peaked velocity distribution on both small and large scales, we now investigate whether it is possible that there are two separate structures in the Spiderweb field. These two structures are possibly in the act of merging with each other, thus creating a larger protocluster structure. We therefore consider the scenario where two haloes, each containing their own group or cluster of galaxies, merge to form a more massive structure. 

One quantative way to tell whether this scenario is feasible is to determine whether the velocity dispersion of $>1000$~km~s$^{-1}$ and  velocity separation between the two groups of $\sim1600$~km~s$^{-1}$ can be reproduced by a merger of two massive dark matter haloes. For this comparison the halo is used that was excluded before due to it being in the process of merging. Based on the number of dark matter particles, these haloes have masses of $2\times10^{13}$ and $6\times10^{13}$~\Msun. 

To trace the velocity evolution, the galaxies belonging to both subhaloes are identified. The locations of the subhaloes are determined by visual inspection and all galaxies within 0.5~Mpc of the centres are assumed to belong to the respective subhaloes. These two groups of galaxies are subsequently traced from $z=2.4$ to $z=1.5$ and their median velocities relative to the velocity of the main dark matter halo are assessed. 

\begin{figure*}
\resizebox{\hsize}{!}{\includegraphics{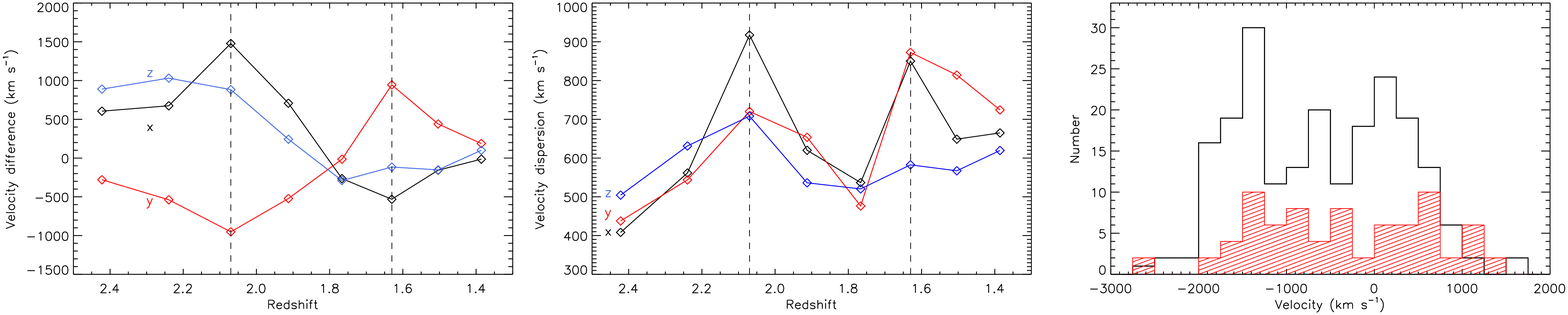}}
\caption{\label{fig:veldifs} Left panel: Velocity difference between two groups of galaxies as a function of redshift. The different colours indicate the different components of the velocity vector, with black, red and blue being the x, y and z component respectively. The dashed lines indicate the approximate redshifts at which mergers take place. At the redshift of the first merger a maximum velocity difference of 1500~km~s$^{-1}$ is seen in the x--component. Middle panel: Evolution of the velocity dispersion of the main halo for all three components as a function of redshift. A maximum is reached at the time of the first merging. Right panel: Velocity distributions of the galaxies in both of the merging haloes at the approximate time of merging. The full Spiderweb velocity distribution is shown as the red hatched histogram and has been scaled up to facilitate comparison. }
\end{figure*}

In the left panel of Fig.~\ref{fig:veldifs} we plot the velocity differences between the two haloes as a function of redshift for each of the three velocity components. The major merger occurs at $z\sim2$ and around this time we see a maximum in the velocity difference for the x--component of approximately 1500~km~s$^{-1}$. This is comparable to the velocity difference found for the Spiderweb. The other velocity components show smaller velocity differences, but this is due to the merger being aligned along the x--axis. The x--component of the velocity therefore represents the velocity obtained in a line-of-sight merger event.

The middle panel of Fig.~\ref{fig:veldifs} shows the evolution of the velocity dispersion of the main halo as a function of redshift. This means that prior to the merger only the velocity dispersion of the most massive halo is considered and during and after the merger both involved haloes are included. We see that at the time of merger the large velocity difference for the x--component results in a sharp increase in the velocity dispersion, reaching a maximum of $\sim900$~km~s$^{-1}$. As with the velocity difference, this is similar to what is found for the Spiderweb protocluster. Another merger occurs at $z\sim1.6$, leading to a similar sharp increase in the velocity dispersion.

The distribution of the x--component of the velocities for both haloes as well as the velocity distribution of the Spiderweb system are shown in the right panel of Fig.~\ref{fig:veldifs}. To facilitate comparison with the distributions shown in Fig.~\ref{fig:velohist}, the velocities have been given with respect to the most massive galaxy in the two haloes. A clear double-peaked structure is seen and in general the resemblance between both distributions is striking. Applying a KS test, we investigate whether the Spiderweb velocity distribution can be drawn from such a velocity distribution and a probability of $\sim40$~per~cent is found. This is a larger probability than found for any of the single haloes investigated above. The hypothesis that the velocity distribution of the Spiderweb is consistent with the distribution caused by the merger of two haloes can therefore not be ruled out. Furthermore, as indicated above, a direct comparison to the distribution of the Spiderweb system has shown that both the velocity difference and the velocity dispersion are only marginally smaller. We therefore consider the merger origin a viable scenario for this protocluster.

Since there is a significant possibility that the Spiderweb system is the result of a merger of two massive haloes, it is now possible to infer something about the properties of the interacting haloes. \citet{venemans2007} calculated the masses of the two groups by assuming that the individual groups are virialized and have virial radii of 0.8 and 1.1~Mpc \citep{kurk2004a}. The velocity dispersions used were 520 and 280~km~s$^{-1}$ as taken from the P00 study. This yielded halo masses of $4\times10^{14}$ and $1\times10^{14}$~\Msun, respectively. With the addition of the confirmed galaxies presented in this work the velocity dispersions become 492 and 417~km~s$^{-1}$ as given in Sect.~\ref{sec:velodis}. This yields that both haloes have a mass of $\sim2.7\times10^{14}$~\Msun. A virialized halo of this mass would have an X-ray luminosity of $\sim2\times10^{43}$~erg~s$^{-1}$ in the restframe 2-10~keV range. Two such haloes in the line of sight are thus consistent with remaining undetected considering the upper limit given by \citet{carilli2002}.

The masses calculated above are significantly larger than the halo masses involved in the simulated merger. However, the masses of the simulated haloes are based on the number of dark matter particles in each halo and may give a more accurate estimate of the actual halo mass. Calculating the masses of the simulated haloes at the time of merger using the velocity dispersion yields masses of $3\times10^{14}$ and $5\times10^{14}$~\Msun, where estimated virial radii of 1~Mpc are used. Using the velocity dispersion in a merger situation can therefore overestimate the total mass by over an order of magnitude. We therefore propose that the subhalo masses stated for the Spiderweb protocluster are to be considered upper mass limits.

In the merger scenario as presented here, it seems odd that the central massive galaxy (\#1) has a velocity placing it between the two velocity peaks seen in Fig.~\ref{fig:velohist}. One would expect the central galaxy to be at the centre of either of the two velocity peaks. However, remember that in Sect.~\ref{sec:velodis} we noted that the redshift of the radio galaxy is highly uncertain due to the large velocity gradients. This could mean that the redshift of the radio galaxy is underestimated. If this is indeed the case, then the radio galaxy would fall exactly within the `red' group.

Finally, it may be argued that this structure is possibly nothing more than a chance superposition of two unrelated structures. It is hard to disprove such a notion, but we must consider that the central radio galaxy is an exceptional object in itself. It has a mass of $\sim10^{12}$~\Msun, hosts a powerful AGN and has an extremely clumpy morphology indicating strong interaction and merging. Furthermore, it has an exceptionally large rotation measure \citep{carilli1997}. If these are indeed separate, unrelated structures it will be difficult to explain why such an exceptional object is found in a relatively low-mass halo. The merger process, however, may be one of the driving forces behind the unique appearance of the central Spiderweb galaxy. This link between the central galaxy and the merger process is further investigated in Sect.~\ref{sec:core}.

\subsubsection{The Spiderweb Galaxy} \label{sec:core}

One of the striking results found in this study, is that the core of the protocluster, which is believed to evolve into a massive cD type galaxy by $z=0$, has a similar velocity distribution as the megaparsec scale protocluster structure. This may purely be the effect of the merger of two protoclusters of galaxies as proposed in Sect.~\ref{sec:merger}, but the large number density of confirmed galaxies near the radio core implies that an additional physical mechanism is at work. In this section we briefly explore a scenario that may explain the dense nature of the core region within the merger scenario.

Although not common to all merging clusters, there have been several reports in the literature of triggered star formation in the constituents of merging clusters \citep{caldwell1997,ferrari2005,hwang2009,ma2010}. Numerical simulations have also shown that it is possible to trigger star formation during cluster merger, either through the tidal gravitational field of the merger \citep{bekki1999} or caused by an increased ram-pressure of the intracluster medium \citep{kronberger2008}. As shown in Fig.~\ref{fig:spiweb}, one of the striking properties of the Spiderweb galaxy is the extended Ly$\alpha$ halo enveloping the system. Much in analogy to what has been found for low-redshift clusters, gravitational interactions and ram pressure may enhance star formation in the galaxies falling through this diffuse halo. This will result in a relatively large number of UV bright galaxies near to the radio core. Naturally, these galaxies reflect the velocity structure of the halo merger as is the case in the protocluster core.

In addition to inducing star formation, the merger may also be responsible for an increase in the AGN activity. The halo merger could lead to a period of more frequent and stronger interactions with the central galaxy. The gas within the radio galaxy would lose its angular momentum through torques induced by these interactions and it would subsequently be funneled towards the central black hole. It is thus possible that a halo merger will trigger a radio luminous phase of the central AGN. If these processes indeed happen, then it is evident that such halo mergers should be considered an important phase of massive galaxy evolution.

We must now ask the question whether selecting galaxy cluster progenitors using HzRGs targets predominantly merging clusters. It is beyond the scope of this paper to answer this question, but other HzRGs have been shown to have relatively large velocity dispersions or are speculated to be part of superstructures \citep[e.g. the 0052 and 0316 protoclusters at $z\sim2.86$ and $z\sim3.13$, respectively][]{venemans2007,maschietto2008}. An extensive study of HzRGs and their protocluster environments is needed to answer this question.

\section{Conclusions} \label{sec:conc}

We have presented results obtained from deep SINFONI data of the core of the protocluster around the radio galaxy PKS~1138-262 at $z\sim2.1$. 

\begin{enumerate}
\item{
We search for emission lines at the locations of the satellite galaxies and find 11 galaxies with one or more emission lines that are consistent with being part of the protocluster. This makes the core region more than an order of magnitude denser than its already overdense large scale environment. Based on the galaxy overdensity of $\sim200$ we estimate that the matter overdensity in the core is $\delta_{\rm m}\sim70$. This is comparable to what is seen in the outskirts of local clusters.
}

\item{
The velocity distribution of the satellite galaxies in the central $120\times120$~kpc$^{2}$ of the protocluster is broad and double-peaked with a velocity dispersion of $1360\pm206$~km~s$^{-1}$. An almost identical velocity distribution was found in a previous study focused on the velocity distribution of the megaparsec scale protocluster structure \citep{pentericci2000,kurk2004b}. This indicates that there is a physical link between the evolution of the central massive galaxy and the larger protocluster structure. Including the results of previous studies, we find that the velocity dispersion of the protocluster is $1013\pm87$~km~s$^{-1}$. This is higher than previous estimates of velocity dispersions in other high-$z$ protoclusters. 
}

\item{
Comparison to the expected field velocity distribution and the peculiar velocity distribution of field environments in the Millennium simulation shows that the protocluster velocity distribution is different at the $4.5\sigma$ and $5\sigma$ significance level, respectively. We therefore conclude that the protocluster is decoupled from the Hubble flow and a dynamically evolved structure.
}

\item{
A comparison to the 25 most massive haloes at $z\sim2$ in the Millennium simulation shows that the majority of the haloes differ from the protocluster at $>2\sigma$ level. However, approximately 20 per cent of these haloes have velocity distributions that differ less than $<2\sigma$ from the Spiderweb system. Based on the velocity distribution we can therefore not exclude the possibility that this is a single virialized massive halo. If virialized, the cluster core will have a mass of $\sim1\times10^{14}$~\Msun~and the total protocluster mass will be $\sim2\times10^{15}$~\Msun. We conclude that the protocluster cannot be virialized since such a total mass would imply an X-ray luminosity of $\sim5\times10^{44}$~erg~s$^{-1}$ and this exceeds the upper limit given in the X-ray study of \citet{carilli2002}. It is possible that only the inner regions have been virialized.
}

\item{
We  investigate an alternative scenario in which the Spiderweb protocluster consists of two galaxy groups that are in the process of merging. A comparison to a massive halo merger at $z\sim2$ in the Millennium simulation shows that velocity differences and dispersions similar to those found in the Spiderweb system can be obtained this way. Furthermore, a KS test shows that the velocity distribution caused by a massive merger of haloes is consistent with the velocity distribution of the Spiderweb protocluster. We conclude that the merger scenario best describes the properties of the protocluster.}

\end{enumerate}

\section*{Acknowledgments}
The authors would like to thank the anonymous referee for all the useful comments that have improved the paper. EK acknowledges funding from Netherlands Organization for Scientific Research (NWO). NAH acknowledges support from STFC and the University of Nottingham Anne McLaren Fellowship. JK thanks the DFG for support via German-Israeli Project Cooperation grant STE1869/1-1.GE625/15-1. This work is based on observations with ESO telescopes at the Paranal Observatories under the programme ID 080.A-0109(A). The Millennium Simulation databases used in this paper and the web application providing online access to them were constructed as part of the activities of the German Astrophysical Virtual Observatory.


\begin{thebibliography}{32}
\expandafter\ifx\csname natexlab\endcsname\relax\def\natexlab#1{#1}\fi

\bibitem[{{Adelberger} {et~al.}(1998){Adelberger}, {Steidel}, {Giavalisco},
  {Dickinson}, {Pettini}, \& {Kellogg}}]{adelberger1998}
{Adelberger} K.~L., {Steidel} C.~C., {Giavalisco} M., {Dickinson} M., {Pettini}
  M., {Kellogg} M., 1998, \apj, 505, 18

\bibitem[{{Beers} {et~al.}(1990){Beers}, {Flynn}, \& {Gebhardt}}]{beers1990}
{Beers} T.~C., {Flynn} K., {Gebhardt} K., 1990, \aj, 100, 32

\bibitem[{{Bekki}(1999)}]{bekki1999}
{Bekki} K., 1999, \apjl, 510, L15

\bibitem[{{Binggeli} {et~al.}(1985){Binggeli}, {Sandage}, \&
  {Tammann}}]{binggeli1985}
{Binggeli} B., {Sandage} A., {Tammann} G.~A., 1985, \aj, 90, 1681

\bibitem[{{Bird} \& {Beers}(1993)}]{bird1993}
{Bird} C.~M., {Beers} T.~C., 1993, \aj, 105, 1596

\bibitem[{{Caldwell} \& {Rose}(1997)}]{caldwell1997}
{Caldwell} N., {Rose} J.~A., 1997, \aj, 113, 492

\bibitem[{{Carilli} {et~al.}(2002){Carilli}, {Harris}, {Pentericci},
  {R{\"o}ttgering}, {Miley}, {Kurk}, \& {van Breugel}}]{carilli2002}
{Carilli} C.~L., {Harris} D.~E., {Pentericci} L., {R{\"o}ttgering} H.~J.~A.,
  {Miley} G.~K., {Kurk} J.~D., {van Breugel} W., 2002, \apj, 567, 781

\bibitem[{{Carilli} {et~al.}(1997){Carilli}, {Roettgering}, {van Ojik},
  {Miley}, \& {van Breugel}}]{carilli1997}
{Carilli} C.~L., {Roettgering} H.~J.~A., {van Ojik} R., {Miley} G.~K., {van
  Breugel} W.~J.~M., 1997, \apjs, 109, 1

\bibitem[{{Croft} {et~al.}(2005){Croft}, {Kurk}, {van Breugel}, {Stanford}, {de
  Vries}, {Pentericci}, \& {R{\"o}ttgering}}]{croft2005}
{Croft} S., {Kurk} J., {van Breugel} W., {Stanford} S.~A., {de Vries} W.,
  {Pentericci} L., {R{\"o}ttgering} H., 2005, \aj, 130, 867

\bibitem[{{De Lucia} \& {Blaizot}(2007)}]{delucia2007}
{De Lucia} G., {Blaizot} J., 2007, \mnras, 375, 2

\bibitem[{{Doherty} {et~al.}(2010){Doherty}, {Tanaka}, {De Breuck}, {Ly},
  {Kodama}, {Kurk}, {Seymour}, {Vernet}, {Stern}, {Venemans}, {Kajisawa}, \&
  {Tanaka}}]{doherty2010}
{Doherty} M. et al., 2010, \aap, 509, A83+

\bibitem[{{Dressler}(1980)}]{dressler1980}
{Dressler} A., 1980, \apj, 236, 351

\bibitem[{{Eisenhauer} {et~al.}(2003){Eisenhauer}, {Abuter}, {Bickert},
  {Biancat-Marchet}, {Bonnet}, {Brynnel}, {Conzelmann}, {Delabre}, {Donaldson},
  {Farinato}, {Fedrigo}, {Genzel}, {Hubin}, {Iserlohe}, {Kasper},
  {Kissler-Patig}, {Monnet}, {Roehrle}, {Schreiber}, {Stroebele}, {Tecza},
  {Thatte}, \& {Weisz}}]{eisenhower2003}
{Eisenhauer} F. et al., 2003, in Presented at the Society
  of Photo-Optical Instrumentation Engineers (SPIE) Conference, Vol. 4841,
  Society of Photo-Optical Instrumentation Engineers (SPIE) Conference Series,
  {M.~Iye \& A.~F.~M.~Moorwood}, ed., pp. 1548--1561

\bibitem[{{Ferrari} {et~al.}(2005){Ferrari}, {Benoist}, {Maurogordato},
  {Cappi}, \& {Slezak}}]{ferrari2005}
{Ferrari} C., {Benoist} C., {Maurogordato} S., {Cappi} A., {Slezak} E., 2005,
  \aap, 430, 19

\bibitem[{{Ford} {et~al.}(1998){Ford}, {Bartko}, {Bely}, {Broadhurst},
  {Burrows}, {Cheng}, {Clampin}, {Crocker}, {Feldman}, {Golimowski}, {Hartig},
  {Illingworth}, {Kimble}, {Lesser}, {Miley}, {Neff}, {Postman}, {Sparks},
  {Tsvetanov}, {White}, {Sullivan}, {Krebs}, {Leviton}, {La Jeunesse},
  {Burmester}, {Fike}, {Johnson}, {Slusher}, {Volmer}, \&
  {Woodruff}}]{ford1998}
{Ford} H.~C. et al., 1998, in Society of Photo-Optical Instrumentation Engineers
  (SPIE) Conference Series, Vol. 3356, Society of Photo-Optical Instrumentation
  Engineers (SPIE) Conference Series, {Bely} P.~Y., {Breckinridge} J.~B., eds.,
  pp. 234--248

\bibitem[{{Galametz} {et~al.}(2010){Galametz}, {Vernet}, {De Breuck}, {Hatch},
  {Miley}, {Kodama}, {Kurk}, {Overzier}, {Rettura}, {Rottgering}, {Seymour},
  {Venemans}, \& {Zirm}}]{galametz2010}
{Galametz} A. et al., 2010, ArXiv e-prints

\bibitem[{{Giavalisco} {et~al.}(1998){Giavalisco}, {Steidel}, {Adelberger},
  {Dickinson}, {Pettini}, \& {Kellogg}}]{giavalisco1998}
{Giavalisco} M., {Steidel} C.~C., {Adelberger} K.~L., {Dickinson} M.~E.,
  {Pettini} M., {Kellogg} M., 1998, \apj, 503, 543

\bibitem[{{Gobat} {et~al.}(2010){Gobat}, {Daddi}, {Onodera}, {Finoguenov},
  {Renzini}, {Arimoto}, {Bouwens}, {Brusa}, {Chary}, {Cimatti}, {Dickinson},
  {Kong}, \& {Mignoli}}]{gobat2010}
{Gobat} R. et al., 2010, ArXiv e-prints

\bibitem[{{Hatch} {et~al.}(2010){Hatch}, {De Breuck}, {Galametz}, {Miley},
  {Overzier}, {R{\"o}ttgering}, {Doherty}, {Kodama}, {Kurk}, {Seymour},
  {Venemans}, {Vernet}, \& {Zirm}}]{hatch2010}
{Hatch} N.~A. et al., 2010, ArXiv e-prints

\bibitem[{{Hatch} {et~al.}(2011){Hatch}, {Kurk}, {Pentericci}, {Venemans},
  {Kuiper}, {Miley}, \& {R{\"o}ttgering}}]{hatch2011}
{Hatch} N.~A., {Kurk} J.~D., {Pentericci} L., {Venemans} B.~P., {Kuiper} E.,
  {Miley} G.~K., {R{\"o}ttgering} H.~J.~A., 2011, ArXiv e-prints

\bibitem[{{Hatch} {et~al.}(2009){Hatch}, {Overzier}, {Kurk}, {Miley},
  {R{\"o}ttgering}, \& {Zirm}}]{hatch2009}
{Hatch} N.~A., {Overzier} R.~A., {Kurk} J.~D., {Miley} G.~K., {R{\"o}ttgering}
  H.~J.~A., {Zirm} A.~W., 2009, \mnras, 395, 114

\bibitem[{{Hatch} {et~al.}(2008){Hatch}, {Overzier}, {R{\"o}ttgering}, {Kurk},
  \& {Miley}}]{hatch2008}
{Hatch} N.~A., {Overzier} R.~A., {R{\"o}ttgering} H.~J.~A., {Kurk} J.~D.,
  {Miley} G.~K., 2008, \mnras, 383, 931

\bibitem[{{Henry} {et~al.}(2010){Henry}, {Salvato}, {Finoguenov}, {Bouche},
  {Brunner}, {Burwitz}, {Buschkamp}, {Egami}, {F{\"o}rster-Schreiber},
  {Fotopoulou}, {Genzel}, {Hasinger}, {Mainieri}, {Rovilos}, \&
  {Szokoly}}]{henry2010}
{Henry} J.~P. et al., 2010, \apj, 725, 615

\bibitem[{{Hwang} \& {Lee}(2009)}]{hwang2009}
{Hwang} H.~S., {Lee} M.~G., 2009, \mnras, 397, 2111

\bibitem[{{Jee} {et~al.}(2009){Jee}, {Rosati}, {Ford}, {Dawson}, {Lidman},
  {Perlmutter}, {Demarco}, {Strazzullo}, {Mullis}, {B{\"o}hringer}, \&
  {Fassbender}}]{jee2009}
{Jee} M.~J. et al., 2009, \apj, 704, 672

\bibitem[{{Knopp} \& {Chambers}(1997)}]{knopp1997}
{Knopp} G.~P., {Chambers} K.~C., 1997, \apjs, 109, 367

\bibitem[{{Kodama} {et~al.}(2007){Kodama}, {Tanaka}, {Kajisawa}, {Kurk},
  {Venemans}, {De Breuck}, {Vernet}, \& {Lidman}}]{kodama2007}
{Kodama} T., {Tanaka} I., {Kajisawa} M., {Kurk} J., {Venemans} B., {De Breuck}
  C., {Vernet} J., {Lidman} C., 2007, \mnras, 377, 1717

\bibitem[{{Kronberger} {et~al.}(2008){Kronberger}, {Kapferer}, {Ferrari},
  {Unterguggenberger}, \& {Schindler}}]{kronberger2008}
{Kronberger} T., {Kapferer} W., {Ferrari} C., {Unterguggenberger} S.,
  {Schindler} S., 2008, \aap, 481, 337

\bibitem[{{Kuiper} {et~al.}(2010){Kuiper}, {Hatch}, {R{\"o}ttgering}, {Miley},
  {Overzier}, {Venemans}, {De Breuck}, {Croft}, {Kajisawa}, {Kodama}, {Kurk},
  {Pentericci}, {Stanford}, {Tanaka}, \& {Zirm}}]{kuiper2010}
{Kuiper} E. et al., 2010, \mnras, 405, 969

\bibitem[{{Kurk} {et~al.}(2000){Kurk}, {R{\"o}ttgering}, {Pentericci}, {Miley},
  {van Breugel}, {Carilli}, {Ford}, {Heckman}, {McCarthy}, \&
  {Moorwood}}]{kurk2000}
{Kurk} J.~D. et al., 2000, \aap, 358, L1

\bibitem[{{Kurk}(2003)}]{kurk2003}
{Kurk} J.~D., 2003, PhD thesis, Leiden University, P.O.~Box 9504, 2300 RA
  Leiden, The Netherlands

\bibitem[{{Kurk} {et~al.}(2004{\natexlab{a}}){Kurk}, {Pentericci}, {Overzier},
  {R{\"o}ttgering}, \& {Miley}}]{kurk2004b}
{Kurk} J.~D., {Pentericci} L., {Overzier} R.~A., {R{\"o}ttgering} H.~J.~A.,
  {Miley} G.~K., 2004{\natexlab{a}}, \aap, 428, 817

\bibitem[{{Kurk} {et~al.}(2004{\natexlab{b}}){Kurk}, {Pentericci},
  {R{\"o}ttgering}, \& {Miley}}]{kurk2004a}
{Kurk} J.~D., {Pentericci} L., {R{\"o}ttgering} H.~J.~A., {Miley} G.~K.,
  2004{\natexlab{b}}, \aap, 428, 793
  
\bibitem[{{Ma} {et~al.}(2010){Ma}, {Ebeling}, {Marshall}, \&
  {Schrabback}}]{ma2010}
{Ma} C., {Ebeling} H., {Marshall} P., {Schrabback} T., 2010, \mnras, 406, 121
  
\bibitem[{{Marinoni} {et~al.}(2005){Marinoni}, {Le F{\`e}vre}, {Meneux},
  {Iovino}, {Pollo}, {Ilbert}, {Zamorani}, {Guzzo}, {Mazure}, {Scaramella},
  {Cappi}, {McCracken}, {Bottini}, {Garilli}, {Le Brun}, {Maccagni}, {Picat},
  {Scodeggio}, {Tresse}, {Vettolani}, {Zanichelli}, {Adami}, {Arnouts},
  {Bardelli}, {Blaizot}, {Bolzonella}, {Charlot}, {Ciliegi}, {Contini},
  {Foucaud}, {Franzetti}, {Gavignaud}, {Marano}, {Mathez}, {Merighi},
  {Paltani}, {Pell{\`o}}, {Pozzetti}, {Radovich}, {Zucca}, {Bondi},
  {Bongiorno}, {Busarello}, {Colombi}, {Cucciati}, {Lamareille}, {Mellier},
  {Merluzzi}, {Ripepi}, \& {Rizzo}}]{marinoni2005}
{Marinoni} C. et al., 2005, \aap, 442, 801  

\bibitem[{{Maschietto} {et~al.}(2008){Maschietto}, {Hatch}, {Venemans},
  {R{\"o}ttgering}, {Miley}, {Overzier}, {Dopita}, {Eisenhardt}, {Kurk},
  {Meurer}, {Pentericci}, {Rosati}, {Stanford}, {van Breugel}, \&
  {Zirm}}]{maschietto2008}
{Maschietto} F. et al., 2008, \mnras, 389, 1223

\bibitem[{{Miley} \& {De Breuck}(2008)}]{miley2008}
{Miley} G., {De Breuck} C., 2008, \aapr, 15, 67

\bibitem[{{Miley} {et~al.}(2006){Miley}, {Overzier}, {Zirm}, {Ford}, {Kurk},
  {Pentericci}, {Blakeslee}, {Franx}, {Illingworth}, {Postman}, {Rosati},
  {R{\"o}ttgering}, {Venemans}, \& {Helder}}]{miley2006}
{Miley} G.~K. et al., 2006, \apjl, 650, L29

\bibitem[{{Navarro} {et~al.}(1996){Navarro}, {Frenk}, \& {White}}]{navarro1996}
{Navarro} J.~F., {Frenk} C.~S., {White} S.~D.~M., 1996, \apj, 462, 563

\bibitem[{{Nesvadba} {et~al.}(2006){Nesvadba}, {Lehnert}, {Eisenhauer},
  {Gilbert}, {Tecza}, \& {Abuter}}]{nesvadba2006}
{Nesvadba} N.~P.~H., {Lehnert} M.~D., {Eisenhauer} F., {Gilbert} A., {Tecza}
  M., {Abuter} R., 2006, \apj, 650, 693

\bibitem[{{Nesvadba} {et~al.}(2008){Nesvadba}, {Lehnert}, {De Breuck},
  {Gilbert}, \& {van Breugel}}]{nesvadba2008}
{Nesvadba} N.~P.~H., {Lehnert} M.~D., {De Breuck} C., {Gilbert} A.~M., {van
  Breugel} W., 2008, \aap, 491, 407

\bibitem[{{Ouchi} {et~al.}(2004){Ouchi}, {Shimasaku}, {Okamura}, {Furusawa},
  {Kashikawa}, {Ota}, {Doi}, {Hamabe}, {Kimura}, {Komiyama}, {Miyazaki},
  {Miyazaki}, {Nakata}, {Sekiguchi}, {Yagi}, \& {Yasuda}}]{ouchi2004}
{Ouchi} M. et al., 2004,
  \apj, 611, 685

\bibitem[{{Overzier} {et~al.}(2008){Overzier}, {Bouwens}, {Cross}, {Venemans},
  {Miley}, {Zirm}, {Ben{\'{\i}}tez}, {Blakeslee}, {Coe}, {Demarco}, {Ford},
  {Homeier}, {Illingworth}, {Kurk}, {Martel}, {Mei}, {Oliveira},
  {R{\"o}ttgering}, {Tsvetanov}, \& {Zheng}}]{overzier2008}
{Overzier} R.~A. et al., 2008, \apj, 673, 143

\bibitem[{{Overzier} {et~al.}(2006){Overzier}, {Miley}, {Bouwens}, {Cross},
  {Zirm}, {Ben{\'{\i}}tez}, {Blakeslee}, {Clampin}, {Demarco}, {Ford},
  {Hartig}, {Illingworth}, {Martel}, {R{\"o}ttgering}, {Venemans}, {Ardila},
  {Bartko}, {Bradley}, {Broadhurst}, {Coe}, {Feldman}, {Franx}, {Golimowski},
  {Goto}, {Gronwall}, {Holden}, {Homeier}, {Infante}, {Kimble}, {Krist}, {Mei},
  {Menanteau}, {Meurer}, {Motta}, {Postman}, {Rosati}, {Sirianni}, {Sparks},
  {Tran}, {Tsvetanov}, {White}, \& {Zheng}}]{overzier2006}
{Overzier} R.~A. et al.,
  2006, \apj, 637, 58
  
\bibitem[{{Owers} {et~al.}(2011){Owers}, {Randall}, {Nulsen}, {Couch}, {David},
  \& {Kempner}}]{owers2011}
{Owers} M.~S., {Randall} S.~W., {Nulsen} P.~E.~J., {Couch} W.~J., {David}
  L.~P., {Kempner} J.~C., 2011, \apj, 728, 27
  
  \bibitem[{{Papovich} {et~al.}(2010){Papovich}, {Momcheva}, {Willmer},
  {Finkelstein}, {Finkelstein}, {Tran}, {Brodwin}, {Dunlop}, {Farrah}, {Khan},
  {Lotz}, {McCarthy}, {McLure}, {Rieke}, {Rudnick}, {Sivanandam}, {Pacaud}, \&
  {Pierre}}]{papovich2010}
{Papovich} C. et al., 2010, \apj, 716, 1503

\bibitem[{{Pascarelle} {et~al.}(1996){Pascarelle}, {Windhorst}, {Driver},
  {Ostrander}, \& {Keel}}]{pascarelle1996}
{Pascarelle} S.~M., {Windhorst} R.~A., {Driver} S.~P., {Ostrander} E.~J.,
  {Keel} W.~C., 1996, \apjl, 456, L21+

\bibitem[{{Pentericci} {et~al.}(2000){Pentericci}, {Kurk}, {R{\"o}ttgering},
  {Miley}, {van Breugel}, {Carilli}, {Ford}, {Heckman}, {McCarthy}, \&
  {Moorwood}}]{pentericci2000}
{Pentericci} L. et al., 2000, \aap, 361, L25

\bibitem[{{Pentericci} {et~al.}(1997){Pentericci}, {Roettgering}, {Miley},
  {Carilli}, \& {McCarthy}}]{pentericci1997}
{Pentericci} L., {Roettgering} H.~J.~A., {Miley} G.~K., {Carilli} C.~L.,
  {McCarthy} P., 1997, \aap, 326, 580
  
\bibitem[{{Poggianti} {et~al.}(2010){Poggianti}, {De Lucia}, {Varela},
  {Aragon-Salamanca}, {Finn}, {Desai}, {von der Linden}, \&
  {White}}]{poggianti2010}
{Poggianti} B.~M., {De Lucia} G., {Varela} J., {Aragon-Salamanca} A., {Finn}
  R., {Desai} V., {von der Linden} A., {White} S.~D.~M., 2010, \mnras, 405, 995
  
\bibitem[{{Ram{\'{\i}}rez} \& {Quintana}(1990)}]{ramirez1990}
{Ram{\'{\i}}rez} A., {Quintana} H., 1990, \rmxaa, 21, 69
  
\bibitem[{{Rocca-Volmerange} {et~al.}(2004){Rocca-Volmerange}, {Le Borgne}, {De
  Breuck}, {Fioc}, \& {Moy}}]{roccavolmerange2004}
{Rocca-Volmerange} B., {Le Borgne} D., {De Breuck} C., {Fioc} M., {Moy} E.,
  2004, \aap, 415, 931
  
 \bibitem[{{Rykoff} {et~al.}(2008){Rykoff}, {Evrard}, {McKay}, {Becker},
  {Johnston}, {Koester}, {Nord}, {Rozo}, {Sheldon}, {Stanek}, \&
  {Wechsler}}]{rykoff2008}
{Rykoff} E.~S. et al., 2008, \mnras, 387, L28

\bibitem[{{Saro} {et~al.}(2009){Saro}, {Borgani}, {Tornatore}, {De Lucia},
  {Dolag}, \& {Murante}}]{saro2009}
{Saro} A., {Borgani} S., {Tornatore} L., {De Lucia} G., {Dolag} K., {Murante}
  G., 2009, \mnras, 392, 795

\bibitem[{{Schmidt} \& {Allen}(2007)}]{schmidt2007}
{Schmidt} R.~W., {Allen} S.~W., 2007, \mnras, 379, 209

\bibitem[{{Seymour} {et~al.}(2007){Seymour}, {Stern}, {De Breuck}, {Vernet},
  {Rettura}, {Dickinson}, {Dey}, {Eisenhardt}, {Fosbury}, {Lacy}, {McCarthy},
  {Miley}, {Rocca-Volmerange}, {R{\"o}ttgering}, {Stanford}, {Teplitz}, {van
  Breugel}, \& {Zirm}}]{seymour2007}
{Seymour} N. et al., 2007, \apjs,
  171, 353
  
\bibitem[{{Small} {et~al.}(1998){Small}, {Ma}, {Sargent}, \&
  {Hamilton}}]{small1998}
{Small} T.~A., {Ma} C., {Sargent} W.~L.~W., {Hamilton} D., 1998, \apj, 492, 45

\bibitem[{{Springel} {et~al.}(2005){Springel}, {White}, {Jenkins}, {Frenk},
  {Yoshida}, {Gao}, {Navarro}, {Thacker}, {Croton}, {Helly}, {Peacock}, {Cole},
  {Thomas}, {Couchman}, {Evrard}, {Colberg}, \& {Pearce}}]{springel2005}
{Springel} V. et al., 2005, \nat, 435, 629

\bibitem[{{Stevens} {et~al.}(2003){Stevens}, {Ivison}, {Dunlop}, {Smail},
  {Percival}, {Hughes}, {R{\"o}ttgering}, {van Breugel}, \&
  {Reuland}}]{stevens2003}
{Stevens} J.~A. et al., 2003, \nat, 425, 264

\bibitem[{{Tanaka} {et~al.}(2010{\natexlab{a}}){Tanaka}, {De Breuck}, {Kurk},
  {Taniguchi}, {Kodama}, {Matsuda}, {Packham}, {Zirm}, {Kajisawa}, {Ichikawa},
  {Seymour}, {Stern}, {Stockton}, {Venemans}, \& {Vernet}}]{itanaka2010}
{Tanaka} I. et al.,
  2010{\natexlab{a}}, ArXiv e-prints

\bibitem[{{Tanaka} {et~al.}(2010){Tanaka}, {Finoguenov}, \&
  {Ueda}}]{tanaka2010}
{Tanaka} M., {Finoguenov} A., {Ueda} Y., 2010, \apjl, 716, L152

\bibitem[{{Venemans} {et~al.}(2005){Venemans}, {R{\"o}ttgering}, {Miley},
  {Kurk}, {De Breuck}, {Overzier}, {van Breugel}, {Carilli}, {Ford}, {Heckman},
  {Pentericci}, \& {McCarthy}}]{venemans2005}
{Venemans} B.~P. et al., 2005, \aap, 431, 793

\bibitem[{{Venemans} {et~al.}(2007){Venemans}, {R{\"o}ttgering}, {Miley}, {van
  Breugel}, {de Breuck}, {Kurk}, {Pentericci}, {Stanford}, {Overzier}, {Croft},
  \& {Ford}}]{venemans2007}
{Venemans} B.~P. et al., 2007, \aap, 461, 823

\bibitem[{{Wilson} {et~al.}(2008){Wilson}, {Muzzin}, {Lacy}, {Yee}, {Surace},
  {Lonsdale}, {Hoekstra}, {Majumdar}, {Gilbank}, \& {Gladders}}]{wilson2008}
{Wilson} G. et al., 2008, in
  Astronomical Society of the Pacific Conference Series, Vol. 381, Infrared
  Diagnostics of Galaxy Evolution, {R.-R.~Chary, H.~I.~Teplitz, \& K.~Sheth},
  ed., pp. 210--+

\bibitem[{{Zirm} {et~al.}(2008){Zirm}, {Stanford}, {Postman}, {Overzier},
  {Blakeslee}, {Rosati}, {Kurk}, {Pentericci}, {Venemans}, {Miley},
  {R{\"o}ttgering}, {Franx}, {van der Wel}, {Demarco}, \& {van
  Breugel}}]{zirm2008}
{Zirm} A.~W. et al., 2008, \apj, 680, 224

\end{thebibliography}

\end{document}